\documentclass{article}

\usepackage{arxiv}

\usepackage[utf8]{inputenc} 
\usepackage[T1]{fontenc}    
\usepackage{hyperref}       
\usepackage{url}            
\usepackage{booktabs}       
\usepackage{amsfonts}       
\usepackage{nicefrac}       
\usepackage{microtype}      
\usepackage{lipsum}		
\usepackage{graphicx}
\usepackage{natbib}
\usepackage{doi}

\usepackage{amsmath,amsfonts}
\usepackage{algorithm}
\usepackage{array}
\usepackage[caption=false,font=normalsize,labelfont=sf,textfont=sf]{subfig}
\usepackage{textcomp}
\usepackage{stfloats}
\usepackage{url}
\usepackage{verbatim}
\usepackage{graphicx}
\usepackage{cite}
\usepackage{booktabs}
\usepackage{makecell}
\usepackage{siunitx} 
\usepackage[dvipsnames]{xcolor}
\usepackage{hyperref}
\hypersetup{
    colorlinks=true,
    linkcolor=NavyBlue,
    filecolor=RedViolet,      
    urlcolor=RedViolet,
    citecolor=NavyBlue,
    }
\usepackage{multirow}

\usepackage{algpseudocode}
\usepackage{tcolorbox}
\tcbuselibrary{skins,breakable}
\usepackage{listings}
\usepackage{xcolor}
\usepackage{fontawesome5}
\definecolor{actorcolor}{RGB}{64, 128, 255}
\definecolor{criticcolor}{RGB}{255, 64, 64}
\definecolor{terminatorcolor}{RGB}{128, 64, 255}
\definecolor{jurorcolor}{RGB}{255, 128, 0}
\definecolor{scenaristcolor}{RGB}{0, 150, 136}
\definecolor{selectorcolor}{RGB}{139, 69, 19}
\definecolor{systemcolor}{RGB}{128, 128, 128}
\definecolor{promptbg}{RGB}{248, 249, 250}
\lstdefinelanguage{JSON}
{
    keywords={true,false,null},
    keywordstyle=\color{blue}\bfseries,
    stringstyle=\color{gray},
    showstringspaces=false,
    morestring=[b]",
    morecomment=[l]{//},
    morecomment=[s]{/*}{*/},
    commentstyle=\color{gray}\itshape,
    basicstyle=\small\ttfamily
}
\newcommand{\iterationheader}[1]{%
    \vspace{1em}
    \noindent\rule{\linewidth}{1pt}
    \vspace{0.2em}
    \begin{center}
        \textbf{\Large Iteration #1}
    \end{center}
    \vspace{0.2em}
    \noindent\rule{\linewidth}{1pt}
    \vspace{0.5em}
}
\newcommand{\scenarioheader}[1]{%
    \vspace{0.5em}
    \begin{center}
        \textbf{\large #1}
    \end{center}
    \vspace{0.3em}
}
\newtcolorbox{systemmetricsbox}{
    colback=white,
    colframe=systemcolor,
    title={{\faCogs\ \textbf{\texttt{System}}}},
    fonttitle=\fontfamily{pcr}\selectfont\bfseries,
    coltitle=white,
    colbacktitle=systemcolor,
    boxrule=1pt,
    arc=3pt,
    left=5pt,
    right=5pt,
    top=5pt,
    bottom=5pt,
    breakable,
    fontupper=\small\ttfamily
}
\newtcolorbox{terminatorpromptbox}{
    colback=promptbg!50,
    colframe=terminatorcolor,
    title={{\faStop\ \textbf{\texttt{LLMTerminator Prompt}}}},
    fonttitle=\fontfamily{pcr}\selectfont\bfseries,
    coltitle=white,
    colbacktitle=terminatorcolor,
    boxrule=1pt,
    arc=3pt,
    left=5pt,
    right=5pt,
    top=5pt,
    bottom=5pt,
    breakable,
    fontupper=\small\ttfamily
}
\newtcolorbox{terminatorresponsebox}{
    colback=terminatorcolor!10,
    colframe=terminatorcolor,
    title={{\faStop\ \textbf{\texttt{LLMTerminator Response}}}},
    fonttitle=\fontfamily{pcr}\selectfont\bfseries,
    coltitle=white,
    colbacktitle=terminatorcolor,
    boxrule=1pt,
    arc=3pt,
    left=5pt,
    right=5pt,
    top=5pt,
    bottom=5pt,
    breakable,
    fontupper=\small\ttfamily
}
\newtcolorbox{jurorpromptbox}{
    colback=promptbg!50,
    colframe=jurorcolor,
    title={{\faGavel\ \textbf{\texttt{LLMJuror Prompt}}}},
    fonttitle=\fontfamily{pcr}\selectfont\bfseries,
    coltitle=white,
    colbacktitle=jurorcolor,
    boxrule=1pt,
    arc=3pt,
    left=5pt,
    right=5pt,
    top=5pt,
    bottom=5pt,
    breakable,
    fontupper=\small\ttfamily
}
\newtcolorbox{jurorresponsebox}{
    colback=jurorcolor!10,
    colframe=jurorcolor,
    title={{\faGavel\ \textbf{\texttt{LLMJuror Response}}}},
    fonttitle=\fontfamily{pcr}\selectfont\bfseries,
    coltitle=white,
    colbacktitle=jurorcolor,
    boxrule=1pt,
    arc=3pt,
    left=5pt,
    right=5pt,
    top=5pt,
    bottom=5pt,
    breakable,
    fontupper=\small\ttfamily
}
\newtcolorbox{actorpromptbox}{
    colback=promptbg!50,
    colframe=actorcolor,
    title={\faRobot\ \textbf{\texttt{LLMActor Prompt}}},
    fonttitle=\fontfamily{pcr}\selectfont\bfseries,
    coltitle=white,
    colbacktitle=actorcolor,
    boxrule=1pt,
    arc=3pt,
    left=5pt,
    right=5pt,
    top=5pt,
    bottom=5pt,
    breakable,
    fontupper=\small\ttfamily
}
\newtcolorbox{actorresponsebox}{
    colback=actorcolor!10,
    colframe=actorcolor,
    title={\faRobot\ \textbf{\texttt{LLMActor Response}}},
    fonttitle=\fontfamily{pcr}\selectfont\bfseries,
    coltitle=white,
    colbacktitle=actorcolor,
    boxrule=1pt,
    arc=3pt,
    left=5pt,
    right=5pt,
    top=5pt,
    bottom=5pt,
    breakable,
    fontupper=\small\ttfamily
}
\newtcolorbox{criticpromptbox}{
    colback=promptbg!50,
    colframe=criticcolor,
    title={{\faSearchMinus\ \textbf{\texttt{LLMCritic Prompt}}}},
    fonttitle=\fontfamily{pcr}\selectfont\bfseries,
    coltitle=white,
    colbacktitle=criticcolor,
    boxrule=1pt,
    arc=3pt,
    left=5pt,
    right=5pt,
    top=5pt,
    bottom=5pt,
    breakable,
    fontupper=\small\ttfamily
}
\newtcolorbox{criticresponsebox}{
    colback=criticcolor!10,
    colframe=criticcolor,
    title={{\faSearchMinus\ \textbf{\texttt{LLMCritic Response}}}},
    fonttitle=\fontfamily{pcr}\selectfont\bfseries,
    coltitle=white,
    colbacktitle=criticcolor,
    boxrule=1pt,
    arc=3pt,
    left=5pt,
    right=5pt,
    top=5pt,
    bottom=5pt,
    breakable,
    fontupper=\small\ttfamily
}
\newtcolorbox{scenaristpromptbox}{
    colback=promptbg!50,
    colframe=scenaristcolor,
    title={{\faFilm\ \textbf{\texttt{LLMScenarist Prompt}}}},
    fonttitle=\fontfamily{pcr}\selectfont\bfseries,
    coltitle=white,
    colbacktitle=scenaristcolor,
    boxrule=1pt,
    arc=3pt,
    left=5pt,
    right=5pt,
    top=5pt,
    bottom=5pt,
    breakable,
    fontupper=\small\ttfamily
}
\newtcolorbox{scenaristresponsebox}{
    colback=scenaristcolor!10,
    colframe=scenaristcolor,
    title={{\faFilm\ \textbf{\texttt{LLMScenarist Response}}}},
    fonttitle=\fontfamily{pcr}\selectfont\bfseries,
    coltitle=white,
    colbacktitle=scenaristcolor,
    boxrule=1pt,
    arc=3pt,
    left=5pt,
    right=5pt,
    top=5pt,
    bottom=5pt,
    breakable,
    fontupper=\small\ttfamily
}
\newtcolorbox{selectorpromptbox}{
    colback=promptbg!50,
    colframe=selectorcolor,
    title={{\faFilter\ \textbf{\texttt{LLMSelector Prompt}}}},
    fonttitle=\fontfamily{pcr}\selectfont\bfseries,
    coltitle=white,
    colbacktitle=selectorcolor,
    boxrule=1pt,
    arc=3pt,
    left=5pt,
    right=5pt,
    top=5pt,
    bottom=5pt,
    breakable,
    fontupper=\small\ttfamily
}
\newtcolorbox{selectorresponsebox}{
    colback=selectorcolor!10,
    colframe=selectorcolor,
    title={{\faFilter\ \textbf{\texttt{LLMSelector Response}}}},
    fonttitle=\fontfamily{pcr}\selectfont\bfseries,
    coltitle=white,
    colbacktitle=selectorcolor,
    boxrule=1pt,
    arc=3pt,
    left=5pt,
    right=5pt,
    top=5pt,
    bottom=5pt,
    breakable,
    fontupper=\small\ttfamily
}

\title{\emph{AgenticControl}: An Automated Control Design Framework Using Large Language Models}

\author{
  \href{https://orcid.org/0009-0008-2436-3099}{\includegraphics[scale=0.06]{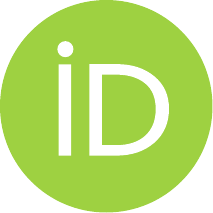}\hspace{1mm}Mohammad Narimani}$^{1,2}$\\
  \texttt{realnarimani@gmail.com} \\
  \And
  \href{https://orcid.org/0000-0002-4131-532X}{\includegraphics[scale=0.06]{orcid.pdf}\hspace{1mm}Seyyed Ali Emami}$^{1,2}$ \thanks{Corresponding author: \texttt{emami@sharif.edu}.}\\
  \texttt{emami@sharif.edu} \\
  \AND
  $^1$Department of Aerospace Engineering \\
  Sharif University of Technology \\
  Tehran, Iran \\
  \AND
  $^2$Artificial Intelligence in Design and Complex Systems (AIDACS) Group \\
  Sharif University of Technology \\
  Tehran, Iran \\
}

\renewcommand{\headeright}{\textit{AgenticControl}: An Automated Control Design Framework Using Large Language Models}
\renewcommand{\undertitle}{}
\renewcommand{\shorttitle}{}

\hypersetup{
pdftitle={AgenticControl: An Automated Control Design Framework Using Large Language Models},
pdfauthor={Mohammad Narimani, Seyyed Ali Emami},
pdfkeywords={Agent-Based Systems, AI-Enabled Robotics, Control Architectures and Programming, Autonomous Agents, Learning from Experience},
}

\begin{document}
\maketitle

\begin{abstract}
Traditional control system design, reliant on expert knowledge and precise models, struggles with complex, nonlinear, or uncertain dynamics. This paper introduces \emph{AgenticControl}, a novel multi-agent framework that automates controller design using coordinated Large Language Model (LLM) agents. Through structured JSON communication, these agents handle tasks including controller selection, scenario design, parameter optimization, performance evaluation, and decision-making. Through an actor-critic optimization approach, the system iteratively improves performance while progressing through scenarios of increasing complexity to ensure robustness under nominal conditions, measurement noise, actuator disturbances, and parametric uncertainties. Key innovations include structured multi-agent collaboration, robust optimization mechanisms, and real-time adaptability via in-context learning. Validated across four diverse control systems, namely, DC Motor Position control, Ball and Beam, Inverted Pendulum, and Double Inverted Pendulum, the framework achieves competitive performance against classical methods. 
Its Full State Feedback solution closely matches Linear Quadratic Regulator (LQR) results, while the designed PID controller significantly outperforming \textsc{Matlab}\textsuperscript{\textregistered}'s PIDTuner, reducing PID tracking error by 55\% through adaptive parameter exploration. A comparative study of five LLM models reveals distinct optimization profiles, with DeepSeek achieving the fastest convergence. This work demonstrates the potential of LLM-driven control design, paving the way for advanced techniques like model predictive control and reinforcement learning. 
\end{abstract}


\section{Introduction}
The integration of Large Language Models (LLMs) into control systems represents a promising frontier in both industrial automation and decision-making processes.
Leveraging their exceptional capabilities in natural language understanding and complex reasoning, LLMs have shown significant potential across diverse applications, including robotics \citep{Liu.2024, Jin.2024.RobotGPT}, autonomous driving \citep{Xu.2024.DriveGPT4}, power electronics \citep{Cui.2024}, and HVAC control \citep{Song.2023}.
Traditional control design methods, such as manual tuning or model-based optimization, often rely heavily on expert knowledge and precise system models, which can be limiting in the face of complex, nonlinear, or uncertain dynamics \citep{Astrom.1995, Franklin.2010}. 
Recent advancements in artificial intelligence, particularly in LLMs, offer a transformative opportunity to automate and enhance this process by combining reasoning capabilities with data-driven insights \citep{Bommasani.2021, Brown.2020}. 

To address these challenges, this paper introduces \emph{AgenticControl}, a novel multi-agent framework that leverages multiple pre-trained LLM agents to systematically design and refine controller parameters across diverse scenarios. The framework integrates multiple specialized LLM agents with computational modules to iteratively propose, evaluate, and optimize controller configurations through simulation and performance assessment. The proposed approach offers several novelties, including:
\begin{enumerate}
    \item \emph{Structured Multi-Agent Collaboration:} 
Our framework introduces a structured multi-agent collaboration that extends existing systems, such as those presented in \citep{He.2024, Jin.2024.RobotGPT, Chen.2024, Guo.2024, Cui.2024}. By integrating multiple LLM agents with computational modules, we ensure both interpretability and precision in the control system design.

    \item \emph{Robust Optimization Mechanisms:} 
We employ unique termination judgment and decision arbitration mechanisms, utilizing dedicated agents to enhance the robustness of the optimization process. This approach distinguishes our framework from single-agent systems or less structured multi-agent designs.

    \item \emph{Real-Time Adaptability:} 
The framework achieves real-time adaptability through the in-context learning capabilities of pre-trained LLMs, eliminating the need for fine-tuning.

    \item \emph{Scalable Workflow:} 
A scalable workflow is implemented using LangGraph, which facilitates efficient coordination among the agents.
\end{enumerate}

\section{Related Works}\label{sec:related_works}
LLMs have shown promise in automating and optimizing control processes, from high-level planning to low-level execution and parameter tuning.
Building on the success of foundation models, many decision-making methods now use LLMs \citep{Sun.2025}.
As categorized by \citet{Song.2023}, these methods generally fall into three categories: (1) fine-tuning LLMs for specific tasks, (2) pairing LLMs with trainable components, and (3) directly utilizing pre-trained LLMs.
Our approach uses pre-trained LLMs, leveraging their In-Context Learning (ICL) abilities for real-time adaptability in various control situations without needing to retrain them.

Within the category of directly utilizing pre-trained LLMs, a significant trend is the use of multi-agent frameworks where LLMs collaborate as specialized agents to address complex control problems. 
For instance, \citet{Chen.2024} proposed an offline Reinforcement Learning (RL) framework where three LLM agents act as state descriptors, action recommenders, and feedback designers to co-design communication and control. Similarly, \citet{Cui.2024} developed a six-agent multi-agent framework for objective-oriented control design in power electronics, with agents handling task management, objective design, model formulation, algorithm development, parameter tuning, and verification.

Theoretical and practical insights further advance LLM-based decision-making. \citet{He.2024} explored hierarchical RL with LLM agents as planner, actor, and reporter, emphasizing exploration for regret minimization. \citet{Park.2024} analyzed LLM agent performance in interactive decision-making using regret. \citet{Guo.2024} introduced ControlAgent, which uses a central agent to distribute tasks to multiple task-specific agents, along with a Python computation agent for iterative parameter refinement. Additionally, \citet{Jin.2024.RobotGPT} developed RobotGPT, which employs a Decision agent, an Evaluation agent, and a Corrector agent to enhance performance. 

Recent studies further highlight the versatility of LLMs in control applications. \citet{Bosio.2024} combined LLMs with evolutionary algorithms to generate interpretable control policies, represented as Python programs. \citet{Tohma.2025} introduced SmartControl, an interactive PID controller design tool where LLM agents translate natural language into numerical targets, optimized via PSO and Differential Evolution, providing comprehensive outputs for academic and simulation use. \citet{Zucchelli.2025} applied fine-tuned LLMs (7-13B parameters) to control simplified space systems, including orbit transfers and powered descent, showing that these LLMs can generate precise multi-dimensional control outputs with less training data and strong generalization. \citet{Zahedifar.2025.LLMController} developed the LLM-Controller, which dynamically tunes the parameters of an adaptive controller to changing conditions through methods like chain-of-thought prompting. 

Earlier works offer further context on LLM capabilities and limitations in control. \citet{Gashi.2024} found that while GPT-4 can interpret control engineering concepts and analytically design controllers for DC motor speed control, it struggles with manual and graphical tuning processes, revealing both potential and limitations. \citet{Tarczewski.2025} developed an LLM-based Tuning Assistant (TA) for PMSM drives, which, despite improving tuning quality, demonstrated that LLMs alone do not guarantee optimality, underscoring the benefit of integrating complementary computational techniques. \citet{Wang.2024} proposed an intelligent industrial automation system integrating the MetaGPT and Langchain frameworks to optimize production lines and reduce carbon emissions. 

\section{Methodology}\label{sec:methodology}

In this section, we present \emph{AgenticControl}, a multi-agent framework that leverages multiple LLM agents to automate the controller design process through a structured optimization framework. 
The system consists of six specialized agents: \texttt{LLMSelector}, \texttt{LLMScenarist}, \texttt{LLMActor}, \texttt{LLMCritic}, \texttt{LLMTerminator}, and \texttt{LLMJuror}, each has a specific task in order to selecting a control law, exploring its design space, and delivering robust control solutions.

\subsection{Framework Overview}

\begin{figure*}[ht]
    \centering
    \includegraphics[width=0.8\linewidth]{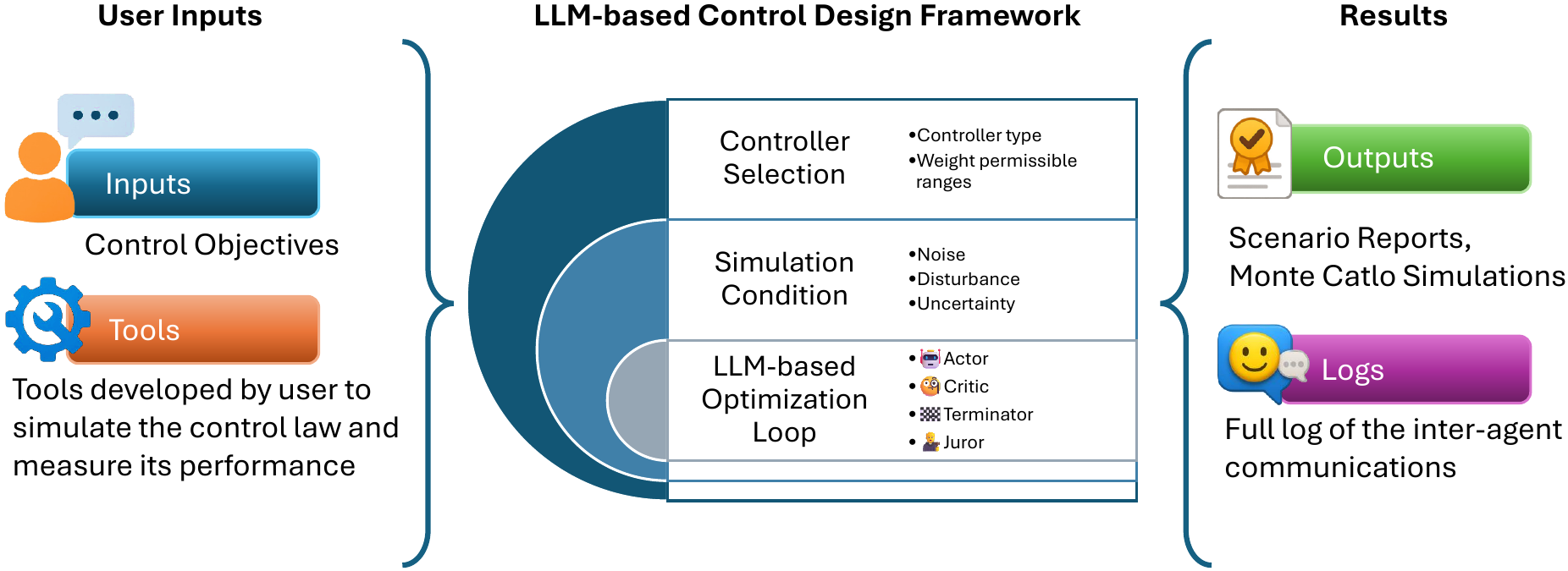}
    \caption{Overview of \emph{AgenticControl}, a multi-agent AI model for automated control system design and optimization. The framework processes user inputs through a structured LLM-based workflow to deliver optimized control solutions across diverse scenarios and controller types.}
    \label{fig:agentic_overview}
\end{figure*}

Fig.~\ref{fig:agentic_overview} illustrates the overall architecture of the proposed method, showing how user inputs are processed through the LLM-based optimization workflow to produce optimized control solutions. The framework provides an end-to-end solution for automated controller design, seamlessly integrating multiple AI agents to handle control system optimization complexity.

\begin{figure}[ht]
    \centering
    \includegraphics[width=0.7\linewidth]{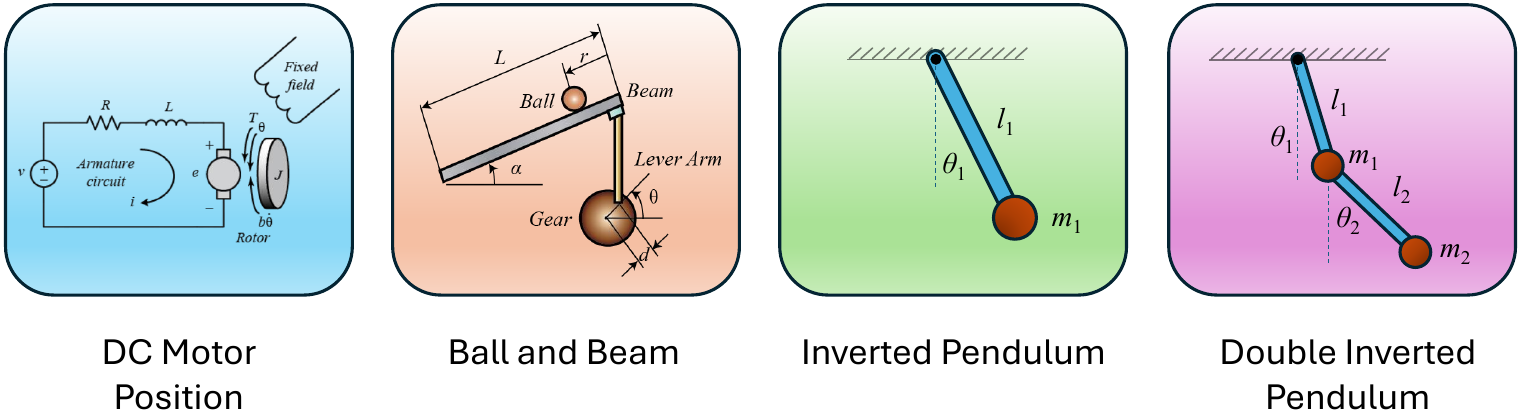}
    \caption{Case study systems for validation: DC Motor Position control, Ball and Beam, Inverted Pendulum, and Double Inverted Pendulum representing diverse control challenges.}
    \label{fig:case_studies}
\end{figure}

The methodology is validated across four representative control systems (Fig.~\ref{fig:case_studies}): DC Motor Position control for precision tracking, Ball and Beam system for nonlinear control, Inverted Pendulum for stability tasks, and Double Inverted Pendulum for complex multi-body dynamics. These systems encompass a broad spectrum from basic linear to advanced nonlinear and unstable configurations.

\subsection{Agent Descriptions}

\subsubsection{LLMSelector}
Chooses control laws sequentially, starting with simpler ones, and sets their parameter search ranges.

\subsubsection{LLMScenarist}
Designs control scenarios of increasing complexity, defining initial conditions, measurement noise, actuator disturbances, and parameter uncertainties.

\subsubsection{LLMActor}
Suggests controller parameters using optimization heuristics and past performance data within specified ranges.

\subsubsection{LLMCritic}
Assesses controller performance, comparing simulation metrics to targets and providing feedback for refinement.

\subsubsection{LLMTerminator}
Decides when to end the optimization loop, evaluating if targets are met or redesign is needed.

\subsubsection{LLMJuror}
Evaluates ambiguous termination cases, deciding on range adjustments, controller changes, or further exploration.

\subsection{Multi-Agent Architecture}

\begin{figure*}[ht]
    \centering
    \includegraphics[width=\linewidth]{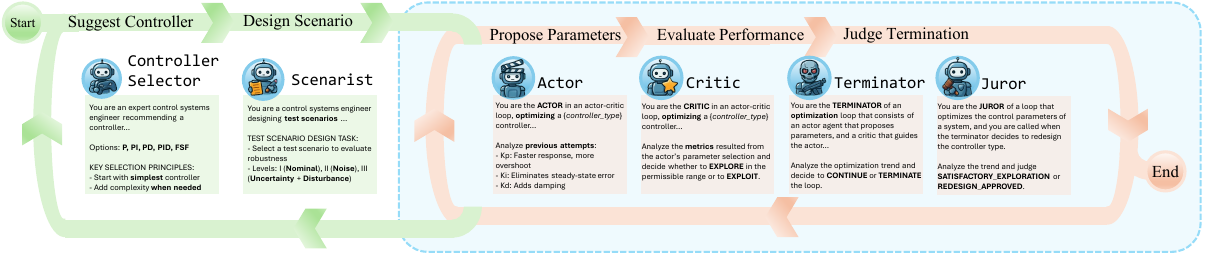}
    \caption{Pipeline of the overall framework.}
    \label{fig:pipeline}
\end{figure*}

\begin{figure}[ht]
    \centering
    \includegraphics[width=0.6\linewidth]{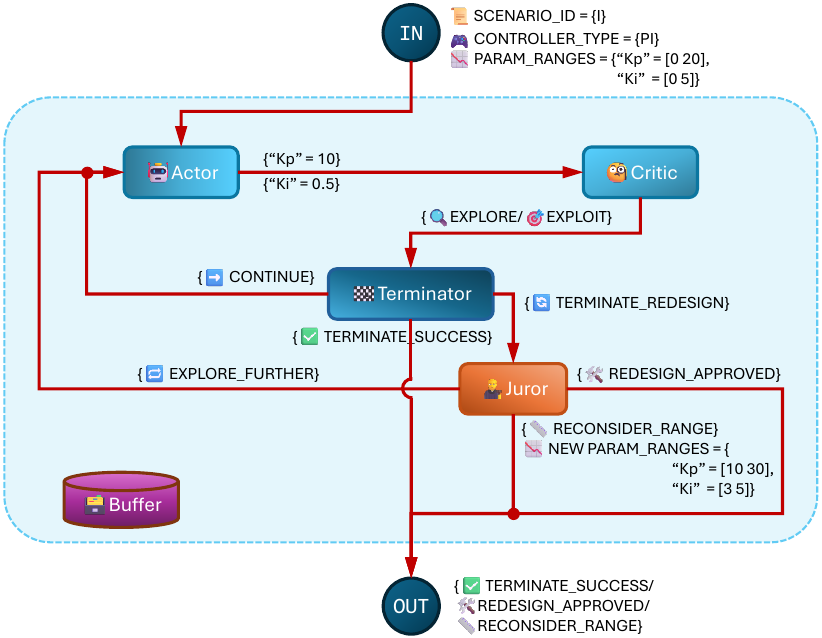}
    \caption{Graph of the inner Actor-critic loop.}
    \label{fig:pipeline_inner}
\end{figure}

The architecture uses six LLM agents communicating via JSON outputs, as shown in Fig.~\ref{fig:pipeline}. 
The inner optimization loop, depicted in Fig.~\ref{fig:pipeline_inner}, adopts an RL-inspired, actor-critic approach, starting with a control type, parameter ranges, and scenario, ending with termination decisions. 
\texttt{LLMSelector} picks the controller and initial ranges, adjustable by \texttt{LLMJuror}; \texttt{LLMScenarist} sets scenario conditions (noise, disturbances, uncertainties); \texttt{LLMActor} proposes parameters; \texttt{LLMCritic} evaluates performance; \texttt{LLMTerminator} decides continuation; and \texttt{LLMJuror} resolves termination disputes. JSON communication ensures efficient workflow coordination.

\subsection{Control Design Workflow}

\subsubsection{System Initialization}
The process begins when the plant dynamics (e.g., DC motor, ball and beam, pendulum) is introduced by the user (Fig.~\ref{fig:agentic_overview}), along with a simulation tool for time-response analysis of the designed control law. For clarity and reproducibility, we focus on a simplified case where the \texttt{LLMSelector} and \texttt{LLMScenarist} are replaced by manual specifications. To this end, we define a predefined list of controllers (P, PI, PD, PID, FSF) and scenarios (nominal, disturbance, uncertainty). Additionally, parameters such as target metrics and maximum iterations are specified manually to facilitate experimental presentation and result reproducibility. These parameters can be determined by dedicated LLMs with default values when needed, but manual specification better serves our validation objectives.

\subsubsection{Iterative Parameter Optimization}
\begin{figure}[ht]
    \centering
    \includegraphics[width=0.6\linewidth]{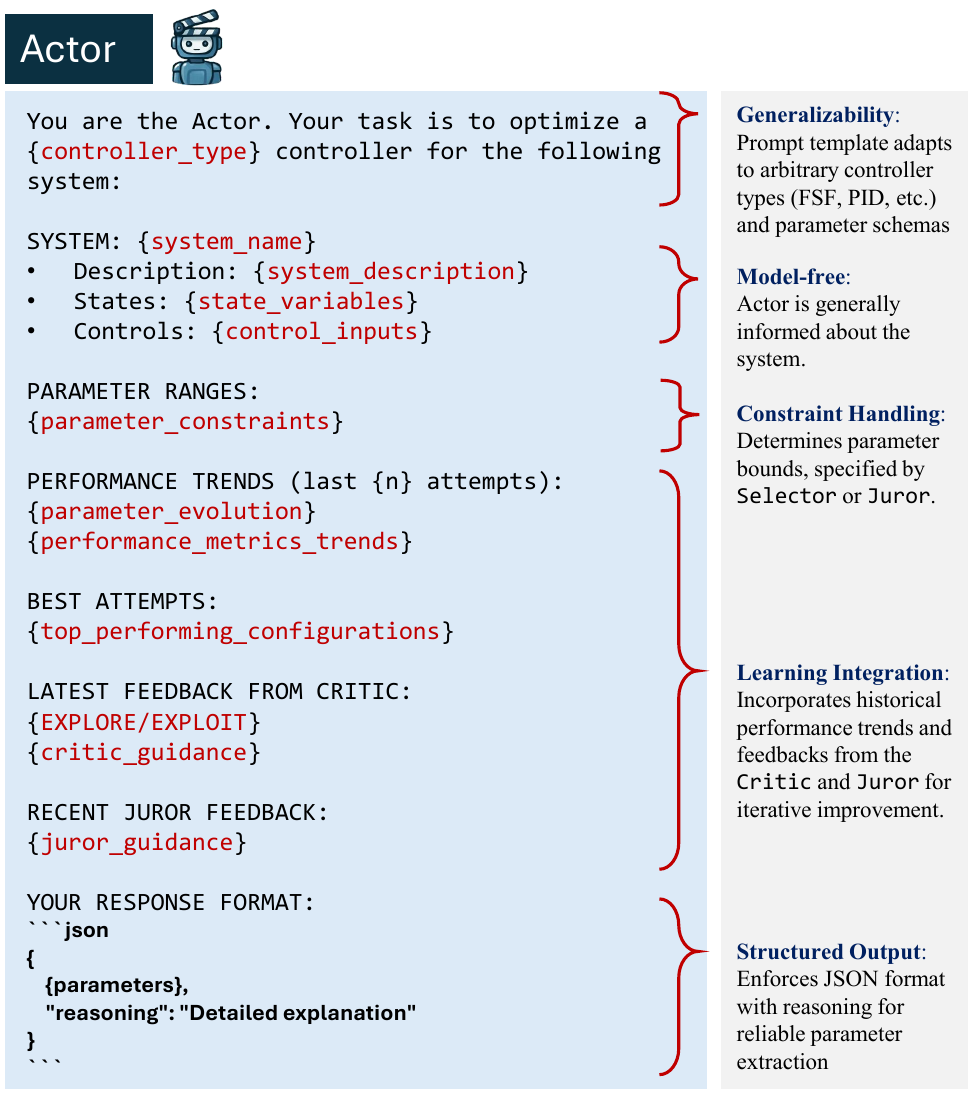}
    \caption{LLMActor prompt structure integrating system descriptions, controller specifications, parameter constraints, and historical performance data for contextually appropriate parameter proposals.}
    \label{fig:llmactor_prompt}
\end{figure}

The core optimization loop consists of five coordinated steps:

\paragraph{Parameter Proposal} The \texttt{LLMActor} generates parameter sets using system description, controller type, parameter ranges, and historical data from a shared buffer. The agent employs control theory principles and past feedback to optimize stability, minimize error, and improve transient response. Fig.~\ref{fig:llmactor_prompt} shows the prompt structure integrating system information and historical data for informed parameter recommendations.

\paragraph{Simulation and Performance Evaluation}

Proposed parameters undergo simulation testing, computing multiple metrics:
\begin{itemize}
    \item \emph{Mean Squared Error}: $\text{MSE} = \frac{1}{N}\sum_{i=1}^{N} e_i^2$
    \item \emph{Settling Time}: Time for response to reach and stay within 2\% of final value
    \item \emph{Overshoot}: Maximum deviation beyond steady-state value
    \item \emph{Rise Time}: Time to reach 90\% of steady-state value
    \item \emph{Zero Crossings}: Number of sign changes in error signal
    \item \emph{Control Effort}: $\int_0^T |u(t)| dt$
    \item \emph{Stability}: Boolean indicator of the closed-loop system stability
\end{itemize}
The \texttt{LLMCritic} then evaluates results, comparing achieved metrics against targets and providing actionable feedback for parameter refinement.

\paragraph{Buffer Update and Termination Decision} The shared buffer updates with latest results, enabling cumulative learning. The \texttt{LLMTerminator} assesses whether performance targets are met or improvements are not sufficient. Subsequently, the \texttt{LLMJuror} provides secondary evaluation for robust decision-making on continuation, parameter range adjustment, or controller switching. It is worth mentioning that the \texttt{LLMJuror} is provided with a statistical analysis of the parameters chosen by \texttt{LLMActor}. Accordingly, the \texttt{LLMJuror} may decide to recommend the  \texttt{LLMActor} to explore further in the current range or modify the search region of the parameters.

\subsubsection{Scenario Progression}
Upon loop completion, successful performance advances to more complex scenarios, while failure triggers controller switching. This ensures only controllers meeting stringent criteria across all scenarios are retained. 

\texttt{LLMScenarist} generates custom scenarios for specific testing needs, while \texttt{LLMSelector} dynamically chooses controllers based on scenario demands and historical performance. These optional agents enhance framework flexibility for specialized applications.

To sum up, AgenticControl leverages structured, agent-driven controller design to systematically select appropriate controller type and refine controller parameters through iterative feedback and scenario-based testing. By integrating LLM capabilities with control theory principles, the framework balances exploration and precision, delivering robust solutions across diverse control challenges. 
In the following section, we evaluate our framework using various case studies to highlight the capabilities of its specialized agents.

\section{Results}\label{sec:results}

This section presents experimental validation of AgenticControl across four case studies with increasing complexity. The evaluation demonstrates progressive controller design, multi-scenario robustification, and comparative LLM model performance.

\subsection{Experimental Setup}

Four representative dynamic systems validate the framework's capabilities. Table~\ref{tab:system_params} summarizes their key characteristics, spanning from linear stable systems to highly nonlinear unstable configurations.

\begin{table*}[htbp]
\centering
\caption{System Parameters and Characteristics for Control Systems}
\label{tab:system_params}
\begin{tabular}{@{}lcccc@{}}
\toprule
\textbf{Parameter} & \textbf{DC Motor} & \textbf{Ball \& Beam} & \textbf{Inverted Pendulum} & \textbf{Double Pendulum} \\
\midrule
System Order       & 2nd & 3rd & 2nd & 4th \\
Linearity          & Linear & Nonlinear & Nonlinear & Nonlinear \\
Stability          & Stable & Non-minimum Phase & Unstable & Unstable \\
\midrule
Primary Dynamics & 
\makecell[t]{$K_m = \SI{0.01}{\newton\meter\per\ampere}$\\ $R = \SI{1}{\ohm}$\\ $L = \SI{0.5}{\henry}$\\ $J = \SI{0.01}{\kilogram\meter\squared}$\\ $b = \SI{0.1}{\newton\meter\second}$\\ $V_{\max} = \SI{24}{\volt}$} & 
\makecell[t]{$g = \SI{9.81}{\meter\per\second\squared}$\\ $r = \SI{0.015}{\meter}$\\ $m = \SI{0.11}{\kilogram}$\\ $J = \SI{1e-5}{\kilogram\meter\squared}$\\ $b = \SI{0.1}{\newton\second\per\meter}$\\ $\theta_{\max} = \SI{5}{\radian\per\second\squared}$} & 
\makecell[t]{$m = \SI{0.1}{\kilogram}$\\ $l = \SI{0.5}{\meter}$\\ $g = \SI{9.81}{\meter\per\second\squared}$\\ $b = \SI{0.1}{\newton\second\per\meter}$\\ $\tau_{\max} = \SI{1}{\newton\meter}$} & 
\makecell[t]{$m_1 = \SI{0.1}{\kilogram}$\\ $m_2 = \SI{0.05}{\kilogram}$\\ $l_1 = \SI{0.5}{\meter}$\\ $l_2 = \SI{0.3}{\meter}$\\ $g = \SI{9.81}{\meter\per\second\squared}$} \\
\bottomrule
\end{tabular}
\end{table*}

\subsection{Progressive Controller Design}

\subsubsection{DC Motor Position Control}
\begin{figure}[ht]
    \centering
    \includegraphics[width=0.65\linewidth]{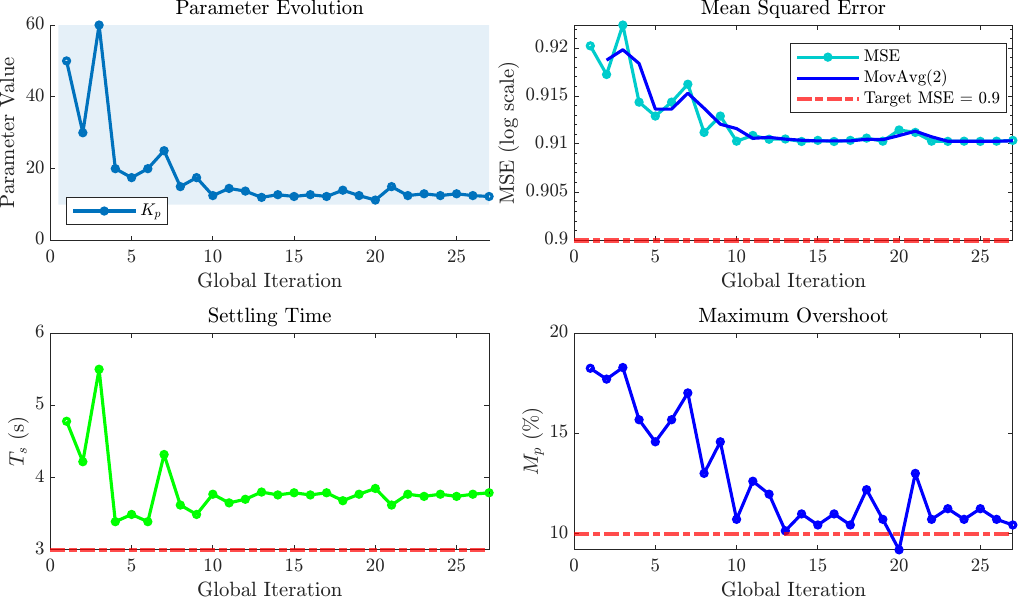}
    \caption{Progressive optimization of the P-controller for DC motor position control, showcasing iterative parameter refinement leading to stable performance with $K_p = 12.75$.}
    \label{fig:dc_motor_progression}
\end{figure}

We start by optimizing a P-controller for the DC Motor Position Control system, iteratively refining the proportional gain. Over 27 iterations, the framework identifies $K_p = 12.75$ as the best parameter, achieving an MSE of 0.9103, settling time of 3.76 s, and overshoot of 10.98\%. The \texttt{LLMTerminator} decides to terminate the optimization, noting parameter convergence and acceptable oscillation levels, despite not fully meeting all target metrics. Fig.~\ref{fig:dc_motor_progression} illustrates the progression of controller performance, guided by systematic parameter adjustments and termination decisions.

\subsubsection{Ball and Beam System Control}
Next, we optimize an FSF controller for the nonlinear Ball and Beam system across three scenarios: nominal conditions, noise with disturbances, and parametric uncertainties.
Iterative optimization yields stable performance, with best parameters achieving MSE $\leq 0.0661$, settling time $T_s \leq 4.18$ s, and overshoot $\leq$ 12.20\% in all scenarios, meeting or approaching targets (MSE $< 0.2$, $T_s < 6.0$ s, overshoot $< 5.0\%$).
Scenario 1, with 14 iterations, achieves the target metrics;  
Scenario 2 requires only 5 iterations; 
Scenario 3, using a total of 47 iterations and two range reconsiderations, obtains $\text{MSE} = 0.0501$.

\begin{figure}[ht]
    \centering
    \includegraphics[width=0.65\linewidth]{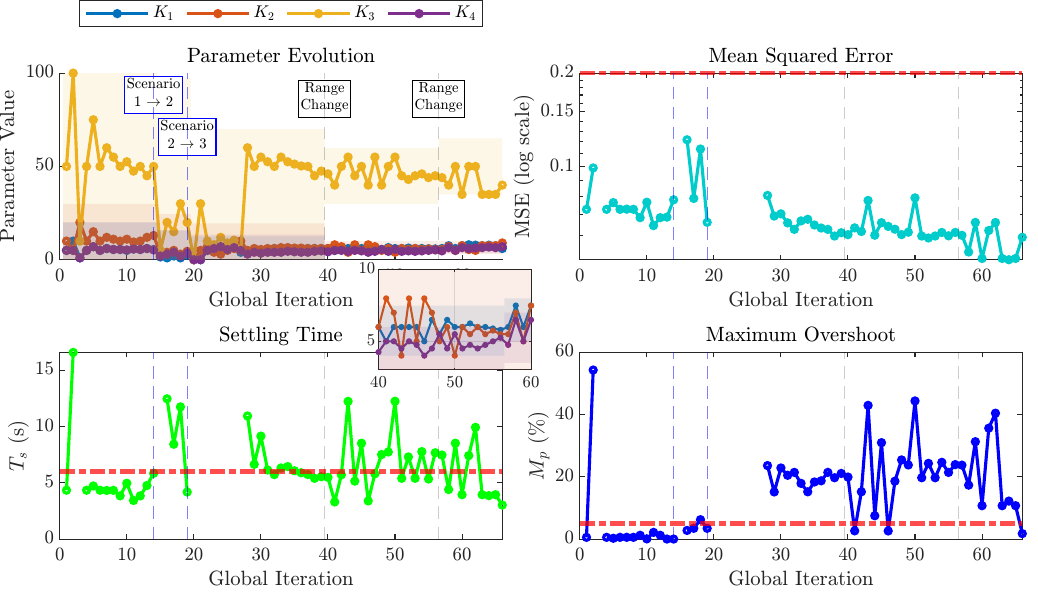}
    \caption{Evolution plots for multi-scenario FSF control design for the Ball and Beam system using AgenticControl.}
    \label{fig:ball_beam_scenarios}
\end{figure}

Finally, Monte Carlo analysis (100 runs) shows our approach outperforms Linear Quadratic Regulator (LQR) ($Q = \mathrm{diag}(100,\ 100,\ 10,\ 10)$, $R = 1$) in MSE, though with higher overshoot.
Fig.~\ref{fig:ball_beam_scenarios} illustrates FSF performance across scenarios, and Table~\ref{tab:ball_beam_results} compares FSF and LQR metrics.

\begin{table}[ht]
\centering
\caption{FSF and LQR Control Results for Ball and Beam Across Scenarios}
\label{tab:ball_beam_results}
\begin{tabular}{ l l c c c c c c c }
\toprule
\textbf{Scenario} & \textbf{Controller} & \textbf{$K_1$} & \textbf{$K_2$} & \textbf{$K_3$} & \textbf{$K_4$} & \textbf{MSE} & \textbf{$T_s$ (s)} & \textbf{$M_p$ (\%)} \\
\midrule
\multirow{2}{*}{Nominal} 
    & AgenticControl & 5.75 & 9.5 & 47.5 & 5.75 & 0.0498 & 3.83 & 11.87 \\
    & LQR & 10.0 & 15.1 & 44.2 & 9.9 & 0.0552 & 3.61 & 0.00 \\
\multirow{2}{*}{Noise + Disturbance} 
    & AgenticControl & 2.5 & 4.5 & 20.0 & 3.5 & 0.0500 & 4.24 & 13.89 \\
    & LQR & — & — & — & — & 0.0552 & 4.06 & 3.45 \\
\multirow{2}{*}{Uncertain Parameters} 
    & AgenticControl & 7.8 & 7.5 & 35.0 & 6.8 & 0.0508 & 3.89 & 12.78 \\
    & LQR & — & — & — & — & 0.0558 & 3.53 & 0.03 \\
\bottomrule
\end{tabular}
\end{table}

\subsection{Multi-Scenario Robustification}

\begin{figure}[ht]
    \centering
    \includegraphics[width=0.65\linewidth]{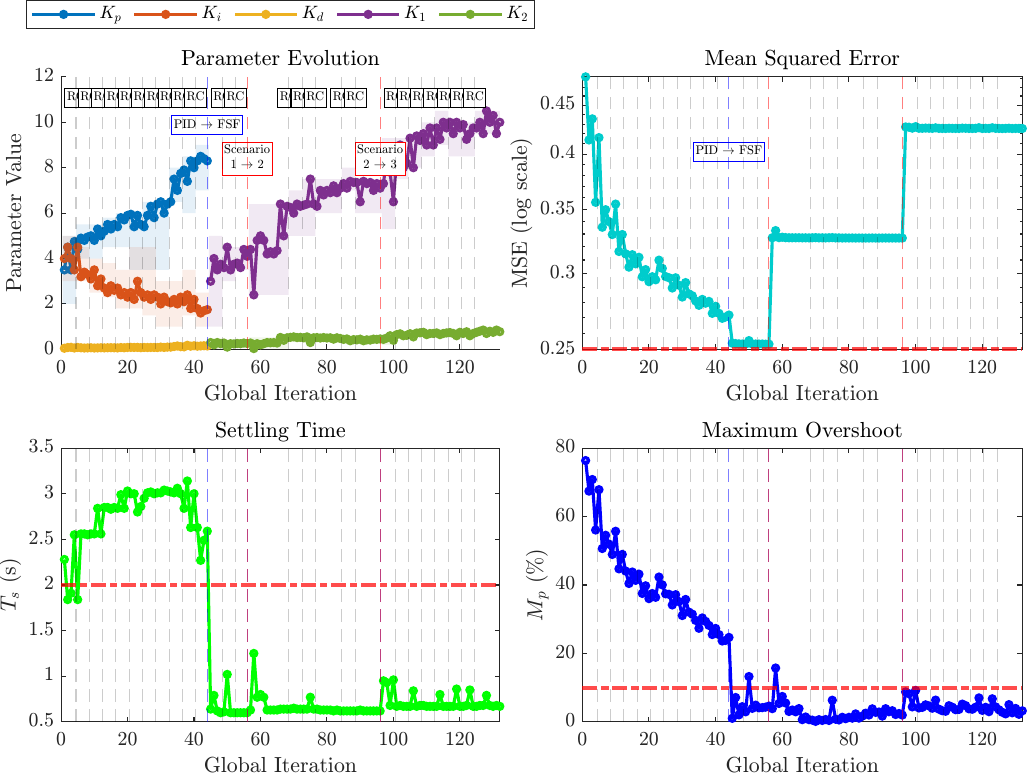}
    \caption{Inverted pendulum optimization progression, showing scenario transitions and \texttt{LLMJuror} interventions for parameter range reconsideration and controller type switching.}
    \label{fig:inverted_pendulum_progression}
\end{figure}

The inverted pendulum case study highlights our framework's ability to design an optimized controller across three scenarios: nominal, measurement noise, and combined uncertainty and disturbance.
To evaluate the effectiveness of \texttt{LLMJuror}'s range reconsideration, we intentionally limited the maximum iterations to 4 per optimization cycle. 
Fig.~\ref{fig:inverted_pendulum_progression} illustrates the design process, emphasizing the adaptive parameter range reconsideration and the transition from PID to FSF after PID reached 10 range reconsiderations in Scenario 1.

The \texttt{LLMJuror}'s \texttt{RECONSIDER\_RANGE} capability adaptively refined parameter ranges, reducing PID's MSE in Scenario 1 from 0.4817 to 0.2694 over 44 iterations with 10 range adjustments (e.g., \texttt{Kp} from [2.0, 5.0] to [7.0, 9.0], \texttt{Ki} to [1.0, 2.5]), though targets (MSE $\leq$ 0.25, $T_s \leq$ 2.0 s, overshoot $\leq$ 10\%) were not fully met.
This prompted a switch to FSF, which achieved near-target performance in Scenario 1 (MSE 0.2531, settling time 0.60 s, overshoot 4.30\%) with only two range reconsiderations. FSF maintained stability in Scenarios 2 and 3, meeting settling time and overshoot targets, though MSE (0.3268 and 0.4254) exceeded the target due to increased complexity.

Table~\ref{tab:inverted_pendulum_results} presents Monte Carlo results for PID and FSF from \emph{AgenticControl}, compared with LQR ($Q = \mathrm{diag}(10,\ 0)$, $R = 0.1$, $K_1=10.50$, $K_2=0.63$) and \textsc{Matlab}\textsuperscript{\textregistered}/Simulink's PIDTuner application ($K_p=1.18$, $K_i=1.94$, $K_d=0.18$).
Our results with FSF tuning outperforms PID across all scenarios, closely matching LQR in Scenario 1 (MSE 0.2531 vs. 0.2529) and surpassing PIDTuner in all scenarios. The framework's adaptive range adjustments and controller switching demonstrate its robustness for complex control tasks.

\begin{table*}[ht]
\centering
\caption{Inverted Pendulum Control Results: PID, FSF, LQR, and PIDTuner Across Scenarios (Monte Carlo, 100 runs)}
\label{tab:inverted_pendulum_results}
\resizebox{\textwidth}{!}{
\begin{tabular}{ l l c c c c }
\toprule
\textbf{Scenario} & \textbf{Controller} & \textbf{Parameters} & \textbf{MSE} & \textbf{$T_s$ (s)} & \textbf{$M_p$ (\%)} \\
\midrule
\multirow{4}{*}{Nominal} & AgenticControl (PID) & $K_p=8.50$, $K_i=1.60$, $K_d=0.15$ & 0.2694 & 2.27 & 23.63 \\
                         & AgenticControl (FSF) & $K_1=4.40$, $K_2=0.28$ & 0.2531 & 0.60 & 4.30 \\
                         & LQR & $K_1=10.50$, $K_2=0.63$ & 0.2529 & 0.59 & 1.44 \\
                         & PIDTuner & $K_p=1.18$, $K_i=1.94$, $K_d=0.18$ & 0.4566 & 2.73 & 64.27 \\
\midrule
\multirow{3}{*}{Noise} & AgenticControl & $K_1=7.30$, $K_2=0.44$ & 0.3268 & 0.62 & 2.51 \\
                       & LQR & — & 0.3268 & 0.62 & 1.32 \\
                       & PIDTuner & — & 0.5897 & 2.77 & 65.29 \\
\midrule
\multirow{3}{*}{Uncertainty \& Disturbance} & AgenticControl & $K_1=10.00$, $K_2=0.80$ & 0.4254 & 0.68 & 2.65 \\
                                & LQR & — & 0.4054 & 0.70 & 4.52 \\
                                & PIDTuner & — & 1.5707 & 4.85 & 79.52 \\
\bottomrule
\end{tabular}
}
\end{table*}

Finally, it's important to note that in this particular case study, the FSF control law functions similarly to a PD controller, but it estimates the pendulum's angular velocity $\dot{\theta}$ via a filter rather than direct measurement. The FSF is expected to perform better because the filter in a PD controller introduces lag, which can degrade its performance.

\subsection{LLM Model Evaluation}

\begin{figure*}[ht]
    \centering
    \includegraphics[width=\linewidth]{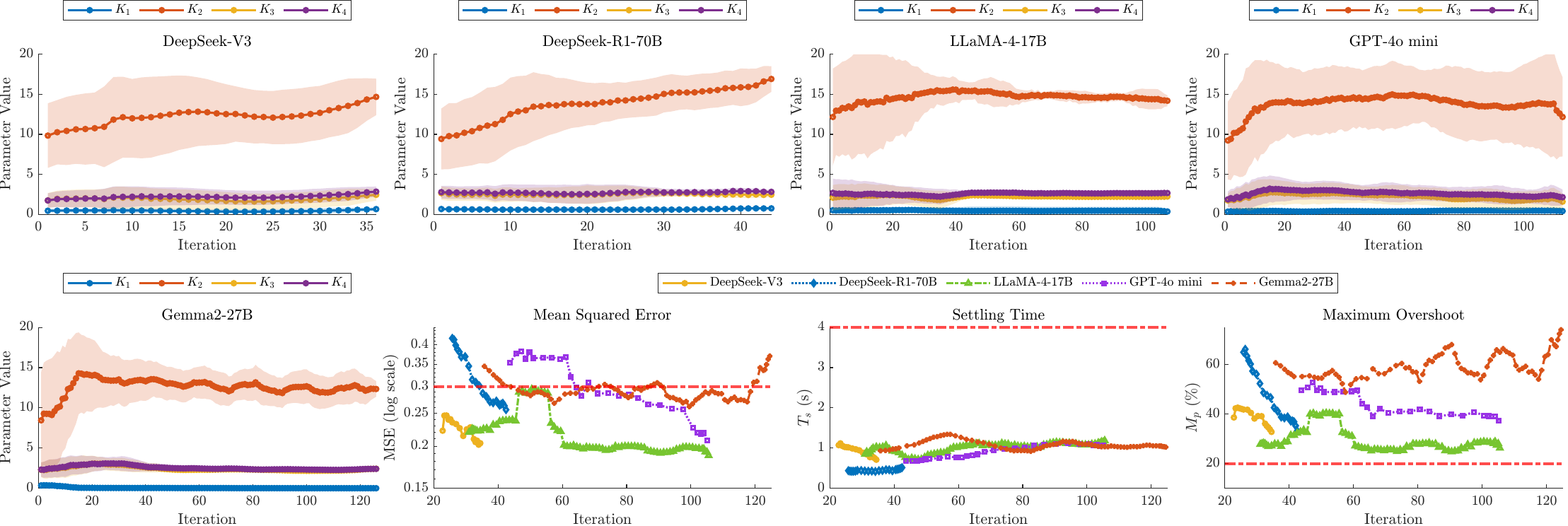}
    \caption{Comparative performance of five LLM models optimizing FSF control for double inverted pendulum, showcasing unique convergence patterns and parameter trends.}
    \label{fig:llm_comparison}
\end{figure*}

In our concluding case study, we assess the effectiveness of various LLM models applied to the double inverted pendulum using FSF control. Fig.~\ref{fig:llm_comparison} illustrates the optimization paths for five models, namely, DeepSeek-V3, DeepSeek-R1-70B, LLaMA-4-17B, GPT-4o mini, and Gemma2-27B, revealing unique convergence and parameter evolution patterns. The plotted curves represent mean values derived from 10 independent runs per model, while the shaded regions denote the standard deviation of the weights, reflecting the exploratory nature of the parameter search. 
DeepSeek models converge around 50 iterations, LLaMA-4-17B and GPT-4o mini near 120, while Gemma2-27B fails to meet performance targets. 

The figure’s lower panel assesses MSE, settling time ($T_s$), and overshoot ($M_p$), with shaded areas omitted for clearer metric visualization. All models struggle to achieve the 20\% overshoot target, but MSE and $T_s$ goals are more attainable. 
DeepSeek-V3 converges fastest, achieving 33\% overshoot by iteration 34, followed by DeepSeek-R1-70B at 42 iterations. LLaMA-4-17B and GPT-4o mini require longer searches, with LLaMA-4-17B reaching 25\% overshoot (12\% lower than GPT-4o mini’s 37\%) and lower metric variance. 
GPT-4o mini, selected for its lower cost and acceptable performance, exhibits the highest variance. 
Gemma2-27B exhibits oscillatory metrics, failing to meet objectives. 

Table~\ref{tab:llm_performance_comparison} presents the best metrics from each model’s 10 runs, benchmarked against the LQR solution (computed using $Q = \text{diag}(1, 1, 10, 10)$, $R = 10$). These metrics are chosen based on the minimum weighted sum of MSE, percent overshoot, and settling time. DeepSeek-V3 achieves the lowest MSE, followed closely by GPT-4o mini. DeepSeek-R1-70B exhibits the smallest overshoot, while LLaMA-4-17B has the shortest settling time but a relatively higher MSE. Gemma2-27B performs poorly, with the highest MSE, likely due to a very low $K_1$ gain. Notably, all LLM models demonstrate superior settling time and control effort compared to the LQR solution, with comparable or even reduced MSE and overshoot.

\begin{table}[ht]
\centering
\caption{Double Inverted Pendulum FSF Control: Best Metrics Achieved by LLM Models vs. LQR Solution Comparison}
\label{tab:llm_performance_comparison}
\begin{tabular}{lcccccccc}
\toprule
\textbf{Method} & \textbf{$K_1$} & \textbf{$K_2$} & \textbf{$K_3$} & \textbf{$K_4$} & \textbf{MSE} & \textbf{$M_p$ (\%)} & \textbf{$T_s$ (s)} & \textbf{Effort} \\
\midrule
DeepSeek-V3 & 0.370 & 12.0 & 0.95 & 1.70 & 0.1328 & 19.16 & 1.14 & 1572.4 \\
DeepSeek-R1-70B & 0.500 & 15.0 & 1.00 & 2.00 & 0.1645 & 13.32 & 1.12 & 1572.1 \\
LLaMA-4-17B & 0.790 & 16.0 & 2.05 & 2.65 & 0.1823 & 18.82 & 0.78 & 1572.7 \\
GPT-4o mini & 0.400 & 12.5 & 0.75 & 1.50 & 0.1395 & 16.53 & 1.10 & 1572.1 \\
Gemma2-27B & 0.005 & 14.8 & 2.00 & 2.50 & 0.1972 & 18.66 & 1.80 & 1572.1 \\
LQR Solution & 0.065 & 11.0 & 1.37 & 2.12 & 0.1500 & 31.83 & 1.71 & 1573.1 \\
\bottomrule
\end{tabular}
\end{table}

These findings highlight distinct LLM performance trade-offs for controller design. DeepSeek-V3 surpasses DeepSeek-R1-70B, with R1’s reasoning reducing parameter search variance but not performance bias; both converge quickly. LLaMA-4-17B delivers consistent, low-variance outcomes despite moderate bias, suiting applications prioritizing reliability. GPT-4o mini, exhibiting high variance and moderate bias, proves less dependable, requiring higher iterations. Gemma2-27B underperforms, with unstable metrics and failure to meet targets, highlighting its inadequacy for complex control tasks.

\section{Conclusion}\label{sec:conclusion}
This paper presented \emph{AgenticControl}, a multi-agent LLM framework for automated controller design, validated across four control systems of increasing complexity. The six-agent architecture successfully automated parameter optimization through structured collaboration, with the \texttt{LLMJuror}'s adaptive range reconsideration proving particularly effective in improving PID performance from MSE 0.4817 to 0.2694 in the inverted pendulum case. Multi-scenario robustification ensured robust performance under nominal conditions, noise, disturbances, and uncertainties, with results matching or exceeding classical methods. LLM model comparison revealed distinct optimization characteristics, with DeepSeek models showing consistent performance.

\emph{AgenticControl} effectively integrates LLM strengths with control engineering needs, delivering a scalable platform that lowers expertise thresholds while preserving accuracy and robustness. The framework's successful validation across linear/nonlinear and stable/unstable systems confirms its versatility and practical applicability. Future research will explore advanced control paradigms (e.g., reinforcement learning, model predictive control, adaptive control) and assess scalability in industrial contexts. This work lays critical groundwork for AI-enhanced control systems, marking a meaningful advancement toward autonomous design solutions for both academic and industrial settings.

\section*{Conflict of interest}
The authors declare no conflict of interest.

\section*{Supplementary Materials}
Additional materials are provided as Supplementary Materials, including full conversation logs of the agentic workflow, and the governing differential equations of the case studies.

\bibliographystyle{unsrtnat}
\bibliography{refs}

\begin{thebibliography}{21}
\providecommand{\natexlab}[1]{#1}
\providecommand{\url}[1]{\texttt{#1}}
\expandafter\ifx\csname urlstyle\endcsname\relax
  \providecommand{\doi}[1]{doi: #1}\else
  \providecommand{\doi}{doi: \begingroup \urlstyle{rm}\Url}\fi

\bibitem[Liu et~al.(2024)Liu, Zhu, Kato, Tsukahara, Kondo, Aoyama, and
  Hasegawa]{Liu.2024}
Haokun Liu, Yaonan Zhu, Kenji Kato, Atsushi Tsukahara, Izumi Kondo, Tadayoshi
  Aoyama, and Yasuhisa Hasegawa.
\newblock Enhancing the llm-based robot manipulation through human-robot
  collaboration.
\newblock \emph{IEEE Robotics and Automation Letters}, 2024.

\bibitem[Jin et~al.(2024)Jin, Li, Shi, Hao, Sun, Zhang, Fang,
  et~al.]{Jin.2024.RobotGPT}
Yixiang Jin, Dingzhe Li, Jun Shi, Peng Hao, Fuchun Sun, Jianwei Zhang, Bin
  Fang, et~al.
\newblock Robotgpt: Robot manipulation learning from chatgpt.
\newblock \emph{IEEE Robotics and Automation Letters}, 9\penalty0 (3):\penalty0
  2543--2550, 2024.

\bibitem[Xu et~al.(2024)Xu, Zhang, Xie, Zhao, Guo, Wong, Li, and
  Zhao]{Xu.2024.DriveGPT4}
Zhenhua Xu, Yujia Zhang, Enze Xie, Zhen Zhao, Yong Guo, Kwan-Yee~K Wong,
  Zhenguo Li, and Hengshuang Zhao.
\newblock Drivegpt4: Interpretable end-to-end autonomous driving via large
  language model.
\newblock \emph{IEEE Robotics and Automation Letters}, 2024.

\bibitem[Cui et~al.(2024)Cui, Liu, Feng, Hui, Ghias, and Zhang]{Cui.2024}
Chenggang Cui, Jiaming Liu, Junkang Feng, Peifeng Hui, Amer~MYM Ghias, and
  Chuanlin Zhang.
\newblock Large language models based multi-agent framework for objective
  oriented control design in power electronics.
\newblock \emph{arXiv preprint arXiv:2406.12628}, 2024.

\bibitem[Song et~al.(2023)Song, Zhang, Zhao, and Bian]{Song.2023}
Lei Song, Chuheng Zhang, Li~Zhao, and Jiang Bian.
\newblock Pre-trained large language models for industrial control.
\newblock \emph{arXiv preprint arXiv:2308.03028}, 2023.

\bibitem[{\AA}str{\"o}m and H{\"a}gglund(1995)]{Astrom.1995}
{Karl Johan} {\AA}str{\"o}m and Tore H{\"a}gglund.
\newblock \emph{PID Controllers: Theory, Design, and Tuning}.
\newblock ISA - The Instrumentation, Systems and Automation Society, 1995.
\newblock ISBN 1-55617-516-7.

\bibitem[Franklin et~al.(2010)Franklin, Powell, Emami-Naeini,
  et~al.]{Franklin.2010}
Gene~F Franklin, J~David Powell, Abbas Emami-Naeini, et~al.
\newblock \emph{Feedback control of dynamic systems}, volume~10.
\newblock Pearson Upper Saddle River, NJ, 2010.

\bibitem[Bommasani et~al.(2021)Bommasani, Hudson, Adeli, Altman, Arora, von
  Arx, Bernstein, Bohg, Bosselut, Brunskill, et~al.]{Bommasani.2021}
Rishi Bommasani, Drew~A Hudson, Ehsan Adeli, Russ Altman, Simran Arora, Sydney
  von Arx, Michael~S Bernstein, Jeannette Bohg, Antoine Bosselut, Emma
  Brunskill, et~al.
\newblock On the opportunities and risks of foundation models.
\newblock \emph{arXiv preprint arXiv:2108.07258}, 2021.

\bibitem[Brown et~al.(2020)Brown, Mann, Ryder, Subbiah, Kaplan, Dhariwal,
  Neelakantan, Shyam, Sastry, Askell, et~al.]{Brown.2020}
Tom Brown, Benjamin Mann, Nick Ryder, Melanie Subbiah, Jared~D Kaplan, Prafulla
  Dhariwal, Arvind Neelakantan, Pranav Shyam, Girish Sastry, Amanda Askell,
  et~al.
\newblock Language models are few-shot learners.
\newblock \emph{Advances in neural information processing systems},
  33:\penalty0 1877--1901, 2020.

\bibitem[He et~al.(2024)He, Chen, Zhang, and Yang]{He.2024}
Jianliang He, Siyu Chen, Fengzhuo Zhang, and Zhuoran Yang.
\newblock From words to actions: Unveiling the theoretical underpinnings of
  llm-driven autonomous systems.
\newblock \emph{arXiv preprint arXiv:2405.19883}, 2024.

\bibitem[Chen et~al.(2024)Chen, Wu, Shen, Ji, Yoshinaga, Ni, Zarakovitis, and
  Zhang]{Chen.2024}
Xianfu Chen, Celimuge Wu, Yi~Shen, Yusheng Ji, Tsutomu Yoshinaga, Qiang Ni,
  Charilaos~C Zarakovitis, and Honggang Zhang.
\newblock Communication and control co-design in 6g: Sequential decision-making
  with llms.
\newblock \emph{IEEE Network}, 2024.

\bibitem[Guo et~al.(2024)Guo, Keivan, Syed, Qin, Zhang, Dullerud, Seiler, and
  Hu]{Guo.2024}
Xingang Guo, Darioush Keivan, Usman Syed, Lianhui Qin, Huan Zhang, Geir
  Dullerud, Peter Seiler, and Bin Hu.
\newblock Controlagent: Automating control system design via novel integration
  of llm agents and domain expertise.
\newblock \emph{arXiv preprint arXiv:2410.19811}, 2024.

\bibitem[Sun et~al.(2025)Sun, Huang, and Pompili]{Sun.2025}
Chuanneng Sun, Songjun Huang, and Dario Pompili.
\newblock Llm-based multi-agent decision-making: Challenges and future
  directions.
\newblock \emph{IEEE Robotics and Automation Letters}, 2025.

\bibitem[Park et~al.(2024)Park, Liu, Ozdaglar, and Zhang]{Park.2024}
Chanwoo Park, Xiangyu Liu, Asuman Ozdaglar, and Kaiqing Zhang.
\newblock Do llm agents have regret? a case study in online learning and games.
\newblock \emph{arXiv preprint arXiv:2403.16843}, 2024.

\bibitem[Bosio and Mueller(2024)]{Bosio.2024}
Carlo Bosio and Mark~W Mueller.
\newblock Synthesizing interpretable control policies through large language
  model guided search.
\newblock \emph{arXiv preprint arXiv:2410.05406}, 2024.

\bibitem[Tohma et~al.(2025)Tohma, Okur, G{\"u}rsoy-Demir, Ayd{\i}n, and
  Yero{\u{g}}lu]{Tohma.2025}
Kadir Tohma, Halil~{\.I}brahim Okur, Handan G{\"u}rsoy-Demir, Merve~Nilay
  Ayd{\i}n, and Celaleddin Yero{\u{g}}lu.
\newblock Smartcontrol: Interactive pid controller design powered by llm agents
  and control system expertise.
\newblock \emph{SoftwareX}, 31:\penalty0 102194, 2025.

\bibitem[Zucchelli et~al.(2025)Zucchelli, Wu, Briden, Hofmann,
  Rodriguez-Fernandez, and Linares]{Zucchelli.2025}
Enrico~M Zucchelli, Di~Wu, Julia Briden, Christian Hofmann, Victor
  Rodriguez-Fernandez, and Richard Linares.
\newblock Fine-tuned language models as space systems controllers.
\newblock \emph{arXiv preprint arXiv:2501.16588}, 2025.

\bibitem[Zahedifar et~al.(2025)Zahedifar, Baghshah, and
  Taheri]{Zahedifar.2025.LLMController}
Rasoul Zahedifar, Mahdieh~Soleymani Baghshah, and Alireza Taheri.
\newblock Llm-controller: Dynamic robot control adaptation using large language
  models.
\newblock \emph{Robotics and Autonomous Systems}, 186:\penalty0 104913, 2025.

\bibitem[Gashi et~al.(2024)Gashi, Chacin, and Tang]{Gashi.2024}
Klison Gashi, Marco Chacin, and Gilbert Tang.
\newblock Electromechanical design and control using a large language model.
\newblock In \emph{2024 12th International Conference on Control, Mechatronics
  and Automation (ICCMA)}, pages 354--360. IEEE, 2024.

\bibitem[Tarczewski et~al.(2025)Tarczewski, Stojic, and
  Dzielinski]{Tarczewski.2025}
Tomasz Tarczewski, Djordje Stojic, and Andrzej Dzielinski.
\newblock Large language model-based tuning assistant for variable speed pmsm
  drive with cascade control structure.
\newblock \emph{Electronics}, 14\penalty0 (2):\penalty0 232, 2025.

\bibitem[Wang and Qin(2024)]{Wang.2024}
Zihao Wang and Huijian Qin.
\newblock Intelligent industrial production process automatic regulation system
  based on llm agents.
\newblock In \emph{2024 5th International Conference on Artificial Intelligence
  and Electromechanical Automation (AIEA)}, pages 133--137. IEEE, 2024.

\end{thebibliography}

\newpage

\section*{Supplementary Materials}

\subsection*{A: Algorithm Description}

The following algorithm outlines the control optimization process implemented in the LangGraph framework. It iteratively selects controllers, designs scenarios, and optimizes parameters using an actor-critic approach, with termination and redesign decisions guided by specialized components.

\begin{algorithm}
\caption{Multi-Agent Controller Design}
\label{alg:multi_agent_opt}
\begin{algorithmic}[1]
\State Initialize system $\mathcal{S}$
\State Initialize controllers $\mathcal{C} = \{\text{P, PI, PD, PID, FSF}\}$
\For{each controller in $\mathcal{C}$}
    \State Select controller type and parameter ranges using \texttt{LLMSelector}
    \For{scenario\_level = 1 to $N_s$}
        \State Select scenario using \texttt{LLMScenarist}
        \State success $\leftarrow$ False
        \For{iteration = 1 to $N_i$}
            \State Select parameters $\theta$ using \texttt{LLMActor}
            \State Simulate $\mathcal{S}$ with $\theta$ and scenario, get metrics
            \State Evaluate metrics using \texttt{LLMCritic}
            \If{\texttt{LLMTerminator} decides success}
                \State success $\leftarrow$ True
                \State \textbf{break}
            \EndIf
        \EndFor
        \If{not success}
            \State Use \texttt{LLMJuror} to decide next action
            \If{\texttt{LLMJuror} decides to reconsider ranges}
                \State Update parameter ranges
                \State Continue with current controller and scenario
            \Else
                \State \textbf{break} \Comment{Move to next controller}
            \EndIf
        \EndIf
    \EndFor
\EndFor
\end{algorithmic}
\end{algorithm}

\newpage

\subsection*{B: Governing Equations of Case Studies}\label{app:governing_equations}

This appendix presents the governing differential equations for the case studies used to validate the AgenticControl framework: DC Motor Position Control, Ball and Beam, Inverted Pendulum, and Double Inverted Pendulum. Parameter values are provided in the original manuscript. These equations describe the system dynamics used for controller optimization.

\paragraph{DC Motor Position Control}
The DC motor position control system models the dynamics of current \( i \), angular velocity \( \omega \), and position \( \theta \). The governing equations are:
\begin{align}
    \dot{i} &= \frac{V - R i - K \omega}{L}, \\
    \dot{\omega} &= \frac{K i - b \omega}{J}, \\
    \dot{\theta} &= \omega,
\end{align}
where \( V \) is the input voltage, and \( J \), \( b \), \( K \), \( R \), and \( L \) are system parameters.

\paragraph{Ball and Beam}
The ball and beam system describes the position \( r \) and velocity \( \dot{r} \) of a ball on a beam with angle \( \alpha \) and angular velocity \( \dot{\alpha} \). The dynamics are:
\begin{align}
    \ddot{r} &= \frac{R^2 m (r \dot{\alpha}^2 - g \sin \alpha)}{m R^2 + J}, \\
    \ddot{\alpha} &= \frac{-2 \dot{\alpha} \dot{r} m r + \theta - g m r \cos \alpha}{m r^2},
\end{align}
where \( \theta \) is the control input (beam angle), and \( g \), \( m \), \( R \), and \( J \) are system parameters.

\paragraph{Inverted Pendulum}
The inverted pendulum models the angle \( \theta \) and angular velocity \( \dot{\theta} \) of a pendulum on a cart. The dynamics are:
\begin{align}
    \ddot{\theta} &= \frac{\tau - L_1 g m_1 \sin \theta - b \dot{\theta}}{L_1^2 m_1},
\end{align}
where \( \tau \) is the input torque, $m_1$ is the pendulum mass, $L_1$ is the rod length, $b$ is the damping coefficient, and $g$ is the gravitational acceleration.

\paragraph{Double Inverted Pendulum}
The double inverted pendulum describes angles \( \theta_1 \), \( \theta_2 \) and angular velocities \( \dot{\theta}_1 \), \( \dot{\theta}_2 \) of two pendulums. Using the Lagrangian approach, the governing equations are derived as
\begin{align}
\ddot{\theta}_1 &= \frac{1}{L_1^2 C} \Biggl[ 
  \begin{aligned}
  &-2 L_1 g m_1 \sin \theta_1 - 2 u + L_1 g m_2 \sin \theta_1 \\
  &+ L_1 g m_2 \sin (\theta_1 - 2 \theta_2) \\
  &+ L_1^2 \dot{\theta}_1^2 m_2 \sin (2 \theta_1 - 2 \theta_2) \\
  &+ 2 L_1 L_2 \dot{\theta}_2^2 m_2 \sin (\theta_1 - \theta_2)
  \end{aligned} \Biggr],\\
\ddot{\theta}_2 &= \frac{1}{L_1 L_2 C} \Biggl[ 
  \begin{aligned}
  &2 L_1^2 \dot{\theta}_1^2 (m_1 + m_2) \sin (\theta_1 - \theta_2)\\
  & - 2 u \cos (\theta_1 - \theta_2) \\
  &- L_1 g (m_1 + m_2) \sin \theta_2 \\
  &+ L_1 g (m_1 + m_2) \sin (2 \theta_1 - \theta_2) \\
  &+ L_1 L_2 \dot{\theta}_2^2 m_2 \sin (2 \theta_1 - 2 \theta_2).
  \end{aligned} \Biggr],
\end{align}
where $ C = 2 m_1 + m_2 - m_2 \cos (2 \theta_1 - 2 \theta_2) $.

\newpage

\subsection*{C: Conversation Logs of Agentic Workflow}
This appendix presents the conversation logs of the agentic workflow for the \emph{AgenticControl} framework, detailing interactions among the LLM agents, each with distinct roles in the optimization process:

\begin{itemize}
    \item \textcolor{actorcolor}{\faRobot\ \textbf{\texttt{LLMActor}}} -- Proposes controller parameter configurations
    \item \textcolor{criticcolor}{\faSearchMinus\ \textbf{\texttt{LLMCritic}}} -- Evaluates proposed configurations and provides feedback
    \item \textcolor{terminatorcolor}{\faStop\ \textbf{\texttt{LLMTerminator}}} -- Decides whether to continue or terminate the optimization process
    \item \textcolor{jurorcolor}{\faGavel\ \textbf{\texttt{LLMJuror}}} -- Makes final decisions on parameter selection
    \item \textcolor{scenaristcolor}{\faFilm\ \textbf{\texttt{LLMScenarist}}} -- Generates simulation scenarios (optional)
    \item \textcolor{selectorcolor}{\faFilter\ \textbf{\texttt{LLMSelector}}} -- Selects controller type (optional)
    \item \textcolor{systemcolor}{\faCogs\ \textbf{\texttt{System}}} -- System messages and simulation results
\end{itemize}

The logs are formatted to show each agent's prompt and response, with performance metrics provided where applicable. For LLM responses, we present their parsed JSON for clarity.

\subsubsection*{Case I: DC Motor}

Here we present the interactions among the LLM agents during the optimization of a Proportional controller for a DC motor position control system.

\begin{systemmetricsbox}
\textbf{=== \faCogs\ SELECTING CONTROLLER ===} \\
\end{systemmetricsbox}

\begin{selectorpromptbox}
System: You are the SELECTOR in a control system optimization framework. Your task is to select the appropriate controller type and initial parameters for the DC motor position control system based on system requirements and optimization goals. \\
\\
\textbf{SYSTEM: DC Motor Position Control} \\
\textbf{DESCRIPTION:} A DC motor position control system with current, angular velocity, and position states. The goal is to control the motor's angular position by applying voltage. The system includes electrical and mechanical dynamics with active control. \\
\\
\textbf{SYSTEM DETAILS:} \\
- Number of states: 3 \\
- State variables: current, omega, theta \\
- Control inputs: voltage \\
\\
\textbf{OPTIMIZATION GOALS:} \\
- Target MSE: 0.9 \\
- Target Settling Time: 3.0s \\
- Target Overshoot: 10\% \\
- Maximum Iterations: 30 \\
\\
\textbf{AVAILABLE CONTROLLER TYPES:} \\
- P: Proportional controller (Kp: [10.0, 60.0]) \\
- PI: Proportional-Integral controller (Kp: [10.0, 60.0], Ki: [0.01, 15.0]) \\
- PD: Proportional-Derivative controller (Kp: [10.0, 60.0], Kd: [0.01, 15.0]) \\
- PID: Proportional-Integral-Derivative controller (Kp: [10.0, 60.0], Ki: [0.01, 20.0], Kd: [0.01, 30.0]) \\
- FSF: Full-State Feedback controller (K1: [0.01, 10.0], K2: [0.01, 100.0], K3: [0.01, 200.0]) \\
\\
\textbf{YOUR TASK:} \\
Select a controller type and initial parameters, considering the system's simplicity and the need for rapid convergence within 30 iterations. Provide reasoning for your choice. \\
\\
\textbf{Required JSON format:} \\
\begin{lstlisting}[language=JSON]
{
    "controller_type": "P|PI|PD|PID|FSF",
    "parameters": {
        "param_name": value,
        ...
    },
    "reasoning": "Explanation of controller selection and initial parameter choices"
}
\end{lstlisting}
\end{selectorpromptbox}

\begin{selectorresponsebox}
\begin{lstlisting}[language=JSON]
{
    "controller_type": "P",
    "parameters": {
        "Kp": 35.0
    },
    "reasoning": "Selected a P controller for its simplicity and suitability for initial exploration in a DC motor system. The initial Kp value is set to the midpoint of the range (35.0) to balance responsiveness and stability, allowing the optimization loop to explore a wide parameter space efficiently."
}
\end{lstlisting}
\end{selectorresponsebox}

\begin{systemmetricsbox}
Selected Controller: P | Params: \{'Kp': 35.0, 'reasoning': 'same as above'\} \\
\textbf{=== \faFilm\ DESIGNING SCENARIO ===} \\
\end{systemmetricsbox}

\begin{scenaristpromptbox}
System: You are the SCENARIST in a control system optimization framework. Your task is to design a test scenario for the DC motor position control system to evaluate the selected controller's performance under specific conditions. \\
\\
\textbf{SYSTEM: DC Motor Position Control} \\
\textbf{DESCRIPTION:} A DC motor position control system with current, angular velocity, and position states. The goal is to control the motor's angular position by applying voltage. The system includes electrical and mechanical dynamics with active control. \\
\\
\textbf{SELECTED CONTROLLER:} \\
- Type: P \\
- Parameters: \{Kp: 35.0\} \\
\\
\textbf{YOUR TASK:} \\
Design a scenario with initial conditions, randomness level, parameter uncertainty, and disturbance level. Ensure the scenario tests the controller's ability to handle nominal conditions for initial evaluation. \\
\\
\textbf{Required JSON format:} \\
\begin{lstlisting}[language=JSON]
{
    "id": "Scenario identifier",
    "randomness_level": float,
    "param_uncertainty": float,
    "initial_condition_range": [float, float],
    "disturbance_level": float,
    "reasoning": "Explanation of scenario design choices"
}
\end{lstlisting}
\end{scenaristpromptbox}

\begin{scenaristresponsebox}
\begin{lstlisting}[language=JSON]
{
    "id": "I",
    "randomness_level": 0.0,
    "param_uncertainty": 0.0,
    "initial_condition_range": [3.14, 3.14],
    "disturbance_level": 0.0,
    "reasoning": "Designed a nominal scenario with fixed initial conditions (theta = pi) and no randomness or disturbances to provide a baseline for evaluating the P controller's performance. This allows clear assessment of the controller's response to a consistent starting position."
}
\end{lstlisting}
\end{scenaristresponsebox}

\scenarioheader{Scenario 1 - DC Motor System}

\begin{systemmetricsbox}
Scenario I: IC Range \protect[$\pi$, $\pi$ \protect] \\
\end{systemmetricsbox}

\iterationheader{1}

\begin{actorpromptbox}
System: You are the ACTOR in a control system optimization framework. Your task is to optimize a P controller for the following system: \\
\\
\textbf{SYSTEM: DC Motor Position Control} \\
\textbf{DESCRIPTION:} A DC motor position control system with current, angular velocity, and position states. The goal is to control the motor's angular position by applying voltage. The system includes electrical and mechanical dynamics with active control. \\
\\
\textbf{SYSTEM DETAILS:} \\
- Number of states: 3 \\
- State variables: current, omega, theta \\
- Control inputs: voltage \\
\\
You are currently at iteration 1 of 30. \\
\\
There are other agents that collaborate with you and provide feedback for you. \\
1- The CRITIC guides you whether to EXPLORE diverse parameter regions or EXPLOIT the current or best-performing parameters. \\
    - For EXPLORE, you need to test a wide range of values within the permissible ranges. In particular, you need to select your parameters near the boundaries in the early iterations, and converge towards possible best-performing point to identify the optimal configuration. The amount of EXPLORATION needs to decrease as we move towards the iterations. \\
    - For EXPLOIT, you need to apply small, precise adjustments to the current or best parameters to enhance stability, reduce error, or improve settling time. \\
    - Based on the current iteration and maximum iterations, move towards EXPLOIT rather than EXPLORE. \\
2- The JUROR analyses your search performance and provides insight about your exploration performance. \\
\\
\textbf{User:} \\
\textbf{CURRENT OPTIMIZATION TASK:} \\
\\
\textbf{P CONTROLLER DETAILS:} \\
- Output feedback controller using current \\
- Kp: Proportional gain (response speed vs overshoot) \\
\\
\textbf{PARAMETER CONSTRAINTS:} \\
- Kp: [10.0000, 60.0000] \\
\\
ALL PARAMETERS MUST BE WITHIN THESE RANGES. \\
\textbf{YOUR RESPONSE FORMAT:} \\
<think> \\
\protect[Analyze the performance trends and determine parameter adjustments. Consider the system dynamics and how each parameter affects the states. Stay within the permissible parameter ranges.\protect] \\
</think> \\
\\
YOU MUST RESPOND WITH THE JSON FORMAT BELOW: \\
\begin{lstlisting}[language=JSON]
{
    "Kp": value,
    "reasoning": "Detailed explanation of parameter choices"
}
\end{lstlisting}
\end{actorpromptbox}

\begin{actorresponsebox}
\begin{lstlisting}[language=JSON]
{
    "Kp": 50.0,
    "reasoning": "Starting with a higher Kp value within the permissible range to explore the system's responsiveness and stability. This choice balances exploration and potential system behavior observation."
}
\end{lstlisting}
\end{actorresponsebox}

\begin{criticpromptbox}
System: You are the CRITIC in a control system optimization framework. Your role is to evaluate controller performance and provide structured feedback for another agent, ACTOR, to improve system performance. \\
\\
Your responsibilities: \\
1. Analyze the current controller performance based on metrics and system response \\
2. Provide strategic guidance for parameter optimization (EXPLORE vs EXPLOIT) \\
3. Give specific, actionable recommendations for parameter adjustments \\
4. Consider trends from previous iterations to guide future improvements \\
\\
Key principles: \\
- EXPLORE: Test wide parameter ranges, especially near boundaries, to find new optimal regions \\
- EXPLOIT: Make small, precise adjustments around current/best parameters to fine-tune performance \\
- Always consider permissible parameter ranges and system stability \\
- Base recommendations on current iteration progress and performance metrics \\
\\
You must respond with valid JSON containing your analysis and recommendations. \\
\textbf{User:} \\
You are currently at iteration 1 of 30. \\
Guide the ACTOR by recommending whether to EXPLORE diverse parameter regions or EXPLOIT the current or best-performing parameters based on performance metrics and iteration progress. \\
\\
\textbf{PERMISSIBLE PARAMETER RANGES:} \\
- Kp: [10.00, 60.00] \\
\\
\textbf{CURRENT CONTROLLER PARAMETERS:} \\
Kp = 50.0000 \\
\\
\textbf{No previous attempts available.} \\
\\
\textbf{No best performance yet.} \\
\\
\textbf{COMPUTED METRICS:} \\
- Mean Squared Error: 0.9203 (Target: 0.9000) \\
- Settling Time: 4.78s (Target: 3.00s) \\
- Maximum Overshoot: 18.2480 rad (Target: 10.0000 rad) \\
- Zero-Crossings: 8 \\
- Control Signal Zero-Crossings: 8 \\
- Control Effort: 9060.5072 \\
- System Stable: Yes \\
\\
\textbf{IMPORTANT: At iteration 1 of 30, the recommended strategy is to EXPLORE.} \\
\\
\textbf{Required JSON format:} \\
\begin{lstlisting}[language=JSON]
{
    "strategy": "EXPLORE or EXPLOIT",
    "result_analysis": "Your concise analysis of the controller performance with key observations",
    "suggested_improvements": [
        "Specific suggestion 1 about parameter adjustments",
        "Specific suggestion 2 about parameter adjustments"
    ]
}
\end{lstlisting}
\end{criticpromptbox}

\begin{criticresponsebox}
\begin{lstlisting}[language=JSON]
{
    "strategy": "EXPLORE",
    "result_analysis": "The controller performance at iteration 1 shows high overshoot and long settling time, indicating an underdamped response. The control effort is also high, suggesting the controller is exerting significant force. Early exploration is needed to find a balance between responsiveness and stability.",
    "suggested_improvements": [
        "Test a lower Kp value, such as 30.00, to reduce overshoot and control effort.",
        "Test a higher Kp value, such as 60.00, to potentially reduce settling time while monitoring for increased overshoot."
    ]
}
\end{lstlisting}
\end{criticresponsebox}

\begin{systemmetricsbox}
\#1/30 | Type:P | Kp:50.000 | MSE:0.9203 | Ts:4.78 | Tr:1.89 | \%OS:18.25 | ZC:8 | CZC:8 | CE:9060.51 | e\_ss:0.02 | isStb:True
\end{systemmetricsbox}

\begin{terminatorpromptbox}
System: You are the TERMINATOR in an ACTOR-CRITIC parameter optimization loop. Your task is to judge whether to continue or terminate the optimization process for controller parameters. Your role is to analyze control system performance metrics, trends, and provide termination decisions in JSON format. \\
\\
Key responsibilities: \\
- Analyze current performance metrics against target metrics \\
- Consider performance trends and improvement patterns \\
- Evaluate controller stability and oscillation characteristics \\
- Make informed decisions about when to terminate optimization \\
- Always respond in valid JSON format \\
\\
Decision types: \\
- TERMINATE\_SUCCESS: Target metrics achieved and controller is stable/non-oscillatory \\
- TERMINATE\_REDESIGN: Progress stalled and sufficient iterations attempted \\
- CONTINUE: Continue optimization process \\
\\
You are currently at iteration 1 of 30. \\
\textbf{User:} \\
Analyze the following control system optimization and decide whether to continue or terminate: \\
\\
\textbf{CURRENT PROGRESS:} \\
- Current iteration: 1 out of 30 maximum iterations \\
- Controller type: P \\
- Minimum required iterations before termination: 6 \\
\\
\textbf{TARGET METRICS:} \\
- MSE target: 0.900000 \\
- Settling time target: 3.00s \\
- Overshoot target: 10.0000 \\
\\
\textbf{CURRENT METRICS:} \\
- MSE: 0.920269 (NOT YET) \\
- Settling time: 4.78s (NOT YET) \\
- Overshoot: 18.2480 (NOT YET) \\
- System stable: Yes \\
- Zero crossings: 8 \\
- Control effort: 9060.5072 \\
- Control zero crossings: 8 \\
\\
\textbf{CRITIC'S STRATEGY:} \\
The critic has suggested to EXPLORE in the last iteration. \\
\\
\textbf{Required JSON format:} \\
\begin{lstlisting}[language=JSON]
{
    "decision": "TERMINATE_SUCCESS|TERMINATE_REDESIGN|CONTINUE",
    "reasoning": "Detailed explanation for your decision",
    "recommendations": "If continuing, specific suggestions for what to focus on next"
}
\end{lstlisting}
\end{terminatorpromptbox}

\begin{terminatorresponsebox}
\begin{lstlisting}[language=JSON]
{
    "decision": "CONTINUE",
    "reasoning": "The current iteration is only 1 out of 30, which is too early to consider termination. The target metrics (MSE, settling time, overshoot) have not been met, and the system is still in the exploration phase. Continuing will allow further exploration of parameter regions to potentially meet the target metrics and improve stability.",
    "recommendations": "Continue with the EXPLORE strategy to test a wider range of parameter values, particularly focusing on reducing overshoot and settling time while maintaining stability."
}
\end{lstlisting}
\end{terminatorresponsebox}

\iterationheader{2}

\begin{actorpromptbox}
System: You are the ACTOR in a control system optimization framework. Your task is to optimize a P controller for the following system: \\
\\
\textbf{SYSTEM: DC Motor Position Control} \\
\textbf{DESCRIPTION:} A DC motor position control system with current, angular velocity, and position states. The goal is to control the motor's angular position by applying voltage. The system includes electrical and mechanical dynamics with active control. \\
\\
\textbf{SYSTEM DETAILS:} \\
- Number of states: 3 \\
- State variables: current, omega, theta \\
- Control inputs: voltage \\
\\
You are currently at iteration 2 of 30. \\
\\
\textbf{RECENT PERFORMANCE HISTORY (1 attempts):} \\
Parameter Trends: \\
- Kp: 50.0000 \\
\\
Performance Trends: \\
- Mse: 0.9203 \\
- Settling Time: 4.78 \\
- Overshoot: 18.2480 \\
- Zero Crossings: 8 \\
- Control Zero Crossings: 8 \\
- Control Effort: 9060.5072 \\
- Stable: Yes \\
\\
\textbf{LATEST FEEDBACK:} \\
- Strategy: EXPLORE \\
- Analysis: The controller performance at iteration 1 shows high overshoot and long settling time, indicating an underdamped response. The control effort is also high, suggesting the controller is exerting significant force. Early exploration is needed to find a balance between responsiveness and stability. \\
- Suggestions: \\
  \textbullet\ Test a lower Kp value, such as 30.00, to reduce overshoot and control effort. \\
  \textbullet\ Test a higher Kp value, such as 60.00, to potentially reduce settling time while monitoring for increased overshoot. \\
\\
\textbf{BEST PERFORMING ATTEMPTS:} \\
Best \#1 (Iteration \#1): \\
- Parameters: Kp=50.0000 \\
- Performance: MSE=0.9203, Settling Time=4.78s, Stable=Yes \\
\\
\textbf{YOUR RESPONSE FORMAT:} \\
<think> \\
\protect[Analyze the performance trends and determine parameter adjustments. Consider the system dynamics and how each parameter affects the states. Stay within the permissible parameter ranges.\protect] \\
</think> \\
\\
YOU MUST RESPOND WITH THE JSON FORMAT BELOW: \\
\begin{lstlisting}[language=JSON]
{
    "Kp": value,
    "reasoning": "Detailed explanation of parameter choices"
}
\end{lstlisting}
\end{actorpromptbox}

\begin{actorresponsebox}
\begin{lstlisting}[language=JSON]
{
    "Kp": 30.00,
    "reasoning": "Reducing Kp to 30.00 to decrease overshoot and control effort while maintaining stability."
}
\end{lstlisting}
\end{actorresponsebox}

\begin{criticpromptbox}
System: You are the CRITIC in a control system optimization framework. Your role is to evaluate controller performance and provide structured feedback for another agent, ACTOR, to improve system performance. \\
\\
You are currently at iteration 2 of 30. \\
\\
\textbf{CURRENT CONTROLLER PARAMETERS:} \\
Kp = 30.0000 \\
\\
\textbf{TREND FROM PREVIOUS 1 RESULTS:} \\
Parameters: \\
Kp: 50.0000 \\
\\
Metrics: \\
mse: 0.9203 \\
settling\_time: 4.7800 \\
overshoot: 18.2480 \\
stable: 1.0000 \\
zero\_crossings: 8.0000 \\
control\_effort: 9060.5072 \\
control\_zero\_crossings: 8.0000 \\
\\
\textbf{BEST PERFORMANCE SO FAR:} \\
Parameters: \\
Kp = 50.0000 \\
\\
Metrics: \\
- Mean Squared Error: 0.9203 \\
- Settling Time: 4.78s \\
- Maximum Overshoot: 18.2480 rad \\
- Zero-Crossings: 8 \\
- Control Effort: 9060.5072 \\
- System Stable: Yes \\
\\
\textbf{COMPUTED METRICS:} \\
- Mean Squared Error: 0.9173 (Target: 0.9000) \\
- Settling Time: 4.22s (Target: 3.00s) \\
- Maximum Overshoot: 17.7074 rad (Target: 10.0000 rad) \\
- Zero-Crossings: 6 \\
- Control Signal Zero-Crossings: 6 \\
- Control Effort: 6560.5487 \\
- System Stable: Yes \\
\\
\textbf{Required JSON format:}
\begin{lstlisting}[language=JSON]
{
    "strategy": "EXPLORE or EXPLOIT",
    "result_analysis": "Your concise analysis of the controller performance with key observations",
    "suggested_improvements": [
        "Specific suggestion 1 about parameter adjustments",
        "Specific suggestion 2 about parameter adjustments"
    ]
}
\end{lstlisting}
\end{criticpromptbox}

\begin{criticresponsebox}
\begin{lstlisting}[language=JSON]
{
    "strategy": "EXPLORE",
    "result_analysis": "The current controller with Kp=30.00 shows improved performance in some metrics but still exceeds targets for MSE, settling time, and overshoot. The system remains stable with reduced control effort. Early exploration is needed to find optimal parameters.",
    "suggested_improvements": [
        "Test Kp=50.00 to confirm previous performance and assess consistency.",
        "Test Kp=60.00 to evaluate boundary conditions and potential improvements.",
        "Test Kp=40.00 to explore intermediate values and observe trends."
    ]
}
\end{lstlisting}
\end{criticresponsebox}

\begin{systemmetricsbox}
\#2/30 | Type:P | Kp:30.000 | MSE:0.9173 | Ts:4.22 | Tr:1.90 | \%OS:17.71 | ZC:6 | CZC:6 | CE:6560.55 | e\_ss:0.00 | isStb:True \\
\end{systemmetricsbox}

\begin{terminatorpromptbox}
System: You are the TERMINATOR in an ACTOR-CRITIC parameter optimization loop. Your task is to judge whether to continue or terminate the optimization process for controller parameters. \\
\\
You are currently at iteration 2 of 30. \\
\\
\textbf{CURRENT PROGRESS:} \\
- Current iteration: 2 out of 30 maximum iterations \\
- Controller type: P \\
- Minimum required iterations before termination: 6 \\
\\
\textbf{TARGET METRICS:} \\
- MSE target: 0.900000 \\
- Settling time target: 3.00s \\
- Overshoot target: 10.0000 \\
\\
\textbf{CURRENT METRICS:} \\
- MSE: 0.917260 (NOT YET) \\
- Settling time: 4.22s (NOT YET) \\
- Overshoot: 17.7074 (NOT YET) \\
- System stable: Yes \\
- Zero crossings: 6 \\
- Control effort: 6560.5487 \\
- Control zero crossings: 6 \\
\\
\textbf{CRITIC'S STRATEGY:} \\
The critic has suggested to EXPLORE in the last iteration. \\
\\
\textbf{TREND FROM PREVIOUS 2 RESULTS:} \\
Parameters: \\
Kp: 50.0000 $\rightarrow$ 30.0000 \\
\\
Metrics: \\
mse: 0.9203 $\rightarrow$ 0.9173 \\
settling\_time: 4.78 $\rightarrow$ 4.22 \\
overshoot: 18.2480 $\rightarrow$ 17.7074 \\
zero\_crossings: 8.0000 $\rightarrow$ 6.0000 \\
control\_effort: 9060.5072 $\rightarrow$ 6560.5487 \\
\\
\textbf{Required JSON format:} \\
\begin{lstlisting}[language=JSON]
{
    "decision": "TERMINATE_SUCCESS|TERMINATE_REDESIGN|CONTINUE",
    "reasoning": "Detailed explanation for your decision",
    "recommendations": "If continuing, specific suggestions for what to focus on next"
}
\end{lstlisting}
\end{terminatorpromptbox}

\begin{terminatorresponsebox}
\begin{lstlisting}[language=JSON]
{
    "decision": "CONTINUE",
    "reasoning": "The current iteration (2) is below the minimum required iterations (6) for termination. The target metrics (MSE: 0.9, settling time: 3.0s, overshoot: 10%) have not been met. While the system is stable, zero crossings (6) and control zero crossings (6) are higher than ideal, indicating some oscillation. The critic's strategy is to EXPLORE, suggesting further parameter exploration is needed. Parameter convergence analysis shows significant changes (40%), indicating parameters are not yet settled. Trends show gradual improvement but not sufficient for termination. Early termination could be premature given the exploration phase and remaining iterations.",
    "recommendations": "Continue with EXPLORE strategy to test a wider range of parameter values, particularly focusing on reducing overshoot and settling time while maintaining stability. Monitor zero crossings to ensure smoother control."
}
\end{lstlisting}
\end{terminatorresponsebox}

\iterationheader{3}

\begin{actorpromptbox}
System: You are the ACTOR in a control system optimization framework. Your task is to optimize a P controller for the following system: \\
\\
\textbf{SYSTEM: DC Motor Position Control} \\
\textbf{DESCRIPTION:} A DC motor position control system with current, angular velocity, and position states. The goal is to control the motor's angular position by applying voltage. The system includes electrical and mechanical dynamics with active control. \\
\\
\textbf{SYSTEM DETAILS:} \\
- Number of states: 3 \\
- State variables: current, omega, theta \\
- Control inputs: voltage \\
\\
You are currently at iteration 3 of 30. \\
\\
\textbf{RECENT PERFORMANCE HISTORY (2 attempts):} \\
Parameter Trends: \\
- Kp: 50.0000 $\rightarrow$ 30.0000 \\
\\
Performance Trends: \\
- Mse: 0.9203 $\rightarrow$ 0.9173 \\
- Settling Time: 4.78 $\rightarrow$ 4.22 \\
- Overshoot: 18.2480 $\rightarrow$ 17.7074 \\
- Zero Crossings: 8 $\rightarrow$ 6 \\
- Control Zero Crossings: 8 $\rightarrow$ 6 \\
- Control Effort: 9060.5072 $\rightarrow$ 6560.5487 \\
- Stable: Yes $\rightarrow$ Yes \\
\\
\textbf{LATEST FEEDBACK:} \\
- Strategy: EXPLORE \\
- Analysis: The current controller with Kp=30.00 shows improved performance in some metrics but still exceeds targets for MSE, settling time, and overshoot. The system remains stable with reduced control effort. Early exploration is needed to find optimal parameters. \\
- Suggestions: \\
  \textbullet\ Test Kp=50.00 to confirm previous performance and assess consistency. \\
  \textbullet\ Test Kp=60.00 to evaluate boundary conditions and potential improvements. \\
  \textbullet\ Test Kp=40.00 to explore intermediate values and observe trends. \\
\\
\textbf{BEST PERFORMING ATTEMPTS:} \\
Best \#1 (Iteration \#2): \\
- Parameters: Kp=30.0000 \\
- Performance: MSE=0.9173, Settling Time=4.22s, Stable=Yes \\
\\
Best \#2 (Iteration \#1): \\
- Parameters: Kp=50.0000 \\
- Performance: MSE=0.9203, Settling Time=4.78s, Stable=Yes \\
\\
\textbf{YOUR RESPONSE FORMAT:} \\
<think> \\
\protect[Analyze the performance trends and determine parameter adjustments. Consider the system dynamics and how each parameter affects the states. Stay within the permissible parameter ranges.\protect] \\
</think> \\
\\
YOU MUST RESPOND WITH THE JSON FORMAT BELOW: \\
\begin{lstlisting}[language=JSON]
{
    "Kp": value,
    "reasoning": "Detailed explanation of parameter choices"
}
\end{lstlisting}
\end{actorpromptbox}

\begin{actorresponsebox}
\begin{lstlisting}[language=JSON]
{
    "Kp": 60.0000,
    "reasoning": "Testing Kp=60.00 to evaluate boundary conditions and potential improvements in MSE, settling time, and overshoot while maintaining stability."
}
\end{lstlisting}
\end{actorresponsebox}

\begin{criticpromptbox}
System: You are the CRITIC in a control system optimization framework. Your role is to evaluate controller performance and provide structured feedback for another agent, ACTOR, to improve system performance. \\
\\
You are currently at iteration 3 of 30. \\
\\
\textbf{CURRENT CONTROLLER PARAMETERS:} \\
Kp = 60.0000 \\
\\
\textbf{TREND FROM PREVIOUS 2 RESULTS:} \\
Parameters: \\
Kp: 50.0000 $\rightarrow$ 30.0000 \\
\\
Metrics: \\
mse: 0.9203 $\rightarrow$ 0.9173 \\
settling\_time: 4.7800 $\rightarrow$ 4.2200 \\
overshoot: 18.2480 $\rightarrow$ 17.7074 \\
zero\_crossings: 8.0000 $\rightarrow$ 6.0000 \\
control\_effort: 9060.5072 $\rightarrow$ 6560.5487 \\
\\
\textbf{BEST PERFORMANCE SO FAR:} \\
Parameters: \\
Kp = 30.0000 \\
\\
Metrics: \\
- Mean Squared Error: 0.9173 \\
- Settling Time: 4.22s \\
- Maximum Overshoot: 17.7074 rad \\
- Zero-Crossings: 6 \\
- Control Effort: 6560.5487 \\
- System Stable: Yes \\
\\
\textbf{COMPUTED METRICS:} \\
- Mean Squared Error: 0.9225 (Target: 0.9000) \\
- Settling Time: 5.50s (Target: 3.00s) \\
- Maximum Overshoot: 18.2861 rad (Target: 10.0000 rad) \\
- Zero-Crossings: 9 \\
- Control Signal Zero-Crossings: 9 \\
- Control Effort: 10439.2154 \\
- System Stable: Yes \\
\\
\textbf{Required JSON format:} \\
\begin{lstlisting}[language=JSON]
{
    "strategy": "EXPLORE or EXPLOIT",
    "result_analysis": "Your concise analysis of the controller performance with key observations",
    "suggested_improvements": [
        "Specific suggestion 1 about parameter adjustments",
        "Specific suggestion 2 about parameter adjustments"
    ]
}
\end{lstlisting}
\end{criticpromptbox}

\begin{criticresponsebox}
\begin{lstlisting}[language=JSON]
{
    "strategy": "EXPLORE",
    "result_analysis": "The current controller performance shows increased MSE, settling time, and overshoot compared to previous iterations, indicating underdamped behavior. The system is not yet optimized, and early exploration is needed to find a better parameter region.",
    "suggested_improvements": [
        "Decrease Kp to 20.00 to explore its effect on reducing overshoot and settling time while monitoring MSE.",
        "Test Kp=10.00 to check if it reduces oscillations and settling time without significantly increasing MSE."
    ]
}
\end{lstlisting}
\end{criticresponsebox}

\begin{systemmetricsbox}
\#3/30 | Type:P | Kp:60.000 | MSE:0.9225 | Ts:5.50 | Tr:1.89 | \%OS:18.29 | ZC:9 | CZC:9 | CE:10439.22 | e\_ss:0.03 | isStb:True
\end{systemmetricsbox}

\begin{terminatorpromptbox}
System: You are the TERMINATOR in an ACTOR-CRITIC parameter optimization loop. Your task is to judge whether to continue or terminate the optimization process for controller parameters. Your role is to analyze control system performance metrics, trends, and provide termination decisions in JSON format. \\
\\
Key responsibilities: \\
- Analyze current performance metrics against target metrics \\
- Consider performance trends and improvement patterns \\
- Evaluate controller stability and oscillation characteristics \\
- Make informed decisions about when to terminate optimization \\
- Always respond in valid JSON format \\
\\
Decision types: \\
- TERMINATE\_SUCCESS: Target metrics achieved and controller is stable/non-oscillatory \\
- TERMINATE\_REDESIGN: Progress stalled and sufficient iterations attempted \\
- CONTINUE: Continue optimization process \\
\\
Critical guidelines: \\
- Prioritize stability and smoothness alongside accuracy \\
- Avoid oscillatory controllers (high zero\_crossings indicates chattering) \\
- Consider ALL metrics, not just MSE, settling time, and overshoot \\
- Require minimum iterations before allowing termination \\
- Require convergence of parameter selection and metrics before allowing successful termination \\
- Always format responses as valid JSON \\
\\
There are other agents that collaborate in this loop. \\
1 - The ACTOR is in charge of selecting controller parameters. You need to give it the chance to EXPLORE different parameters, as well as letting it EXPLOIT around its best-performing parameters. \\
2 - The CRITIC guides whether to EXPLORE diverse parameter regions or EXPLOIT the current or best-performing parameters. \\
    - For EXPLORE, we need to test a wide range of values within the permissible ranges, particularly near the boundaries in the early iterations, to identify new optimal configurations. \\
    - For EXPLOIT, we need to observe the effect of applying small, precise adjustments to the current or best parameters to enhance stability, reduce error, or improve settling time. \\
    - Based on the current iteration and maximum iterations, we need to move towards EXPLOIT rather than EXPLORE. \\
\\
\textbf{User:} \\
Analyze the following control system optimization and decide whether to continue or terminate: \\
\\
\textbf{SYSTEM DESCRIPTION:} \\
A DC motor position control system with current, angular velocity, and position states. The goal is to control the motor's angular position by applying voltage. The system includes electrical and mechanical dynamics with active control. \\
\\
\textbf{CURRENT PROGRESS:} \\
- Current iteration: 3 out of 30 maximum iterations \\
- Controller type: P \\
- Minimum required iterations before termination: 6 \\
\\
\textbf{TARGET METRICS:} \\
- MSE target: 0.900000 \\
- Settling time target: 3.00s \\
- Overshoot target: 10.0000 \\
\\
\textbf{CURRENT METRICS:} \\
- MSE: 0.922464 (NOT YET) \\
- Settling time: 5.50s (NOT YET) \\
- Overshoot: 18.2861 (NOT YET) \\
- System stable: Yes \\
- Zero crossings: 9 (Lower is better - indicates fewer oscillations) \\
- Control effort: 10439.2154 (Lower is better - indicates efficient control) \\
- Control zero crossings: 9 (Lower is better - indicates smoother control) \\
\\
\textbf{CRITIC'S STRATEGY:} \\
The critic has suggested to EXPLORE in the last iteration. \\
- `EXPLORE' indicates searching for better parameter regions, termination might be premature \\
- `EXPLOIT' indicates fine-tuning around promising area, termination could be considered if metrics are satisfactory \\
- `UNKNOWN' means no strategy information available; default to CONTINUE \\
\\
\textbf{TREND FROM PREVIOUS 3 RESULTS:} \\
Parameters: \\
Kp: 50.0000 $\rightarrow$ 30.0000 $\rightarrow$ 60.0000 \\
\\
Metrics: \\
mse: 0.9203 $\rightarrow$ 0.9173 $\rightarrow$ 0.9225 \\
settling\_time: 4.78 $\rightarrow$ 4.22 $\rightarrow$ 5.50 \\
overshoot: 18.2480 $\rightarrow$ 17.7074 $\rightarrow$ 18.2861 \\
stable: 1.0000 $\rightarrow$ 1.0000 $\rightarrow$ 1.0000 \\
zero\_crossings: 8.0000 $\rightarrow$ 6.0000 $\rightarrow$ 9.0000 \\
control\_effort: 9060.5072 $\rightarrow$ 6560.5487 $\rightarrow$ 10439.2154 \\
control\_zero\_crossings: 8.0000 $\rightarrow$ 6.0000 $\rightarrow$ 9.0000 \\
\\
\textbf{IMPROVEMENT ANALYSIS:} \\
\begin{lstlisting}[language=JSON]
{
  "mse_change": -0.2385,
  "settling_time_change": -15.0628,
  "overshoot_change": -0.2085,
  "iterations_analyzed": 3
}
\end{lstlisting}
\textbf{PARAMETER CONVERGENCE ANALYSIS:} \\
\begin{lstlisting}[language=JSON]
{
  "parameter_changes": {
    "Kp": 70.0
  },
  "max_change_percent": 70.0,
  "converged": false,
  "iterations_analyzed": 3
}
\end{lstlisting}
\textbf{DECISION CRITERIA:} \\
- If current\_iteration < 6, always CONTINUE \\
- If critic\_strategy == `UNKNOWN', CONTINUE \\
- If critic\_strategy == `EXPLORE', CONTINUE \\
- If critic\_strategy == `EXPLOIT': \\
  * If all target metrics are met (MSE <= 0.9, settling\_time <= 3.0, overshoot <= 10), zero\_crossings <= 5, AND parameter convergence is achieved (max\_change\_percent <= 5\% in PARAMETER CONVERGENCE ANALYSIS), consider TERMINATE\_SUCCESS \\
  * If metrics are not meeting targets or zero\_crossings > 5 or parameters are not converged (max\_change\_percent > 20\% in PARAMETER CONVERGENCE ANALYSIS), and no significant improvement in recent iterations (e.g., MSE improvement < 5\% over last few iterations), consider TERMINATE\_REDESIGN \\
  * Otherwise, CONTINUE \\
- Always prioritize stability (stable == True), low oscillations (zero\_crossings <= 5), and smooth control (control\_zero\_crossings <= 10) \\
\\
\textbf{Provide your decision in the following JSON format:} \\
\begin{lstlisting}[language=JSON]
{
    "decision": "TERMINATE_SUCCESS|TERMINATE_REDESIGN|CONTINUE",
    "reasoning": "Detailed explanation for your decision",
    "recommendations": "If continuing, specific suggestions for what to focus on next"
}
\end{lstlisting}
\end{terminatorpromptbox}

\begin{terminatorresponsebox}
\begin{lstlisting}[language=JSON]
{
    "decision": "CONTINUE",
    "reasoning": "The current iteration (3) is below the minimum required iterations (6) for termination. The system is stable, but none of the target metrics (MSE, settling time, overshoot) have been met. The parameters are still changing significantly (70% change in Kp), indicating that the system has not yet converged. The critic's strategy is to EXPLORE, suggesting that termination would be premature. While there have been some improvements in MSE and overshoot, settling time has increased, and zero crossings are higher than desired, indicating potential oscillations. Continuing the optimization process will allow further exploration and potential improvement in all metrics.",
    "recommendations": "Continue with the optimization process, focusing on exploring different parameter regions to improve MSE, reduce settling time, lower overshoot, and minimize oscillations (zero crossings)."
}
\end{lstlisting}
\end{terminatorresponsebox}

\iterationheader{4-27}

\begin{systemmetricsbox}
\#4/30 | Type:P | Kp:20.000 | MSE:0.9144 | Ts:3.39 | Tr:1.94 | \%OS:15.68 | ZC:5 | CZC:5 | CE:5382.83 | e\_ss:0.00 | isStb:True \\
\#5/30 | Type:P | Kp:17.500 | MSE:0.9129 | Ts:3.49 | Tr:1.96 | \%OS:14.58 | ZC:5 | CZC:5 | CE:5060.60 | e\_ss:0.00 | isStb:True \\
\#6/30 | Type:P | Kp:20.000 | MSE:0.9144 | Ts:3.39 | Tr:1.94 | \%OS:15.68 | ZC:5 | CZC:5 | CE:5382.83 | e\_ss:0.00 | isStb:True \\
\#7/30 | Type:P | Kp:25.000 | MSE:0.9162 | Ts:4.32 | Tr:1.91 | \%OS:17.02 | ZC:6 | CZC:6 | CE:5980.34 | e\_ss:0.00 | isStb:True \\
\#8/30 | Type:P | Kp:15.000 | MSE:0.9112 | Ts:3.62 | Tr:2.00 | \%OS:13.01 | ZC:4 | CZC:4 | CE:4712.27 | e\_ss:0.00 | isStb:True \\
\#9/30 | Type:P | Kp:17.500 | MSE:0.9129 | Ts:3.49 | Tr:1.96 | \%OS:14.58 | ZC:5 | CZC:5 | CE:5060.60 | e\_ss:0.00 | isStb:True \\
\#10/30 | Type:P | Kp:12.500 | MSE:0.9103 | Ts:3.77 | Tr:2.09 | \%OS:10.71 | ZC:4 | CZC:4 | CE:4322.42 | e\_ss:0.00 | isStb:True \\
\#11/30 | Type:P | Kp:14.500 | MSE:0.9109 | Ts:3.65 | Tr:2.02 | \%OS:12.61 | ZC:4 | CZC:4 | CE:4638.11 | e\_ss:0.00 | isStb:True \\
\#12/30 | Type:P | Kp:13.750 | MSE:0.9105 | Ts:3.70 | Tr:2.04 | \%OS:11.97 | ZC:4 | CZC:4 | CE:4523.34 | e\_ss:0.00 | isStb:True \\
\#13/30 | Type:P | Kp:12.000 | MSE:0.9105 | Ts:3.80 | Tr:2.12 | \%OS:10.14 | ZC:4 | CZC:4 | CE:4238.35 | e\_ss:0.00 | isStb:True \\
\#14/30 | Type:P | Kp:12.750 | MSE:0.9103 | Ts:3.76 | Tr:2.08 | \%OS:10.98 | ZC:4 | CZC:4 | CE:4363.64 | e\_ss:0.00 | isStb:True \\
\#15/30 | Type:P | Kp:12.250 | MSE:0.9104 | Ts:3.79 | Tr:2.10 | \%OS:10.43 | ZC:4 | CZC:4 | CE:4280.65 | e\_ss:0.00 | isStb:True \\
\#16/30 | Type:P | Kp:12.750 | MSE:0.9103 | Ts:3.76 | Tr:2.08 | \%OS:10.98 | ZC:4 | CZC:4 | CE:4363.64 | e\_ss:0.00 | isStb:True \\
\#17/30 | Type:P | Kp:12.250 | MSE:0.9104 | Ts:3.79 | Tr:2.10 | \%OS:10.43 | ZC:4 | CZC:4 | CE:4280.65 | e\_ss:0.00 | isStb:True \\
\#18/30 | Type:P | Kp:14.000 | MSE:0.9106 | Ts:3.68 | Tr:2.03 | \%OS:12.19 | ZC:4 | CZC:4 | CE:4562.04 | e\_ss:0.00 | isStb:True \\
\#19/30 | Type:P | Kp:12.500 | MSE:0.9103 | Ts:3.77 | Tr:2.09 | \%OS:10.71 | ZC:4 | CZC:4 | CE:4322.42 | e\_ss:0.00 | isStb:True \\
\#20/30 | Type:P | Kp:11.250 | MSE:0.9115 | Ts:3.85 | Tr:2.17 | \%OS:9.19 | ZC:3 | CZC:3 | CE:4108.04 | e\_ss:0.00 | isStb:True \\
\#21/30 | Type:P | Kp:15.000 | MSE:0.9112 | Ts:3.62 | Tr:2.00 | \%OS:13.01 | ZC:4 | CZC:4 | CE:4712.27 | e\_ss:0.00 | isStb:True \\
\#22/30 | Type:P | Kp:12.500 | MSE:0.9103 | Ts:3.77 | Tr:2.09 | \%OS:10.71 | ZC:4 | CZC:4 | CE:4322.42 | e\_ss:0.00 | isStb:True \\
\#23/30 | Type:P | Kp:13.000 | MSE:0.9103 | Ts:3.74 | Tr:2.07 | \%OS:11.25 | ZC:4 | CZC:4 | CE:4404.30 | e\_ss:0.00 | isStb:True \\
\#24/30 | Type:P | Kp:12.500 | MSE:0.9103 | Ts:3.77 | Tr:2.09 | \%OS:10.71 | ZC:4 | CZC:4 | CE:4322.42 | e\_ss:0.00 | isStb:True \\
\#25/30 | Type:P | Kp:13.000 | MSE:0.9103 | Ts:3.74 | Tr:2.07 | \%OS:11.25 | ZC:4 | CZC:4 | CE:4404.30 | e\_ss:0.00 | isStb:True \\
\#26/30 | Type:P | Kp:12.500 | MSE:0.9103 | Ts:3.77 | Tr:2.09 | \%OS:10.71 | ZC:4 | CZC:4 | CE:4322.42 | e\_ss:0.00 | isStb:True \\
\#27/30 | Type:P | Kp:12.250 | MSE:0.9104 | Ts:3.79 | Tr:2.10 | \%OS:10.43 | ZC:4 | CZC:4 | CE:4280.65 | e\_ss:0.00 | isStb:True \\
\end{systemmetricsbox}

\begin{terminatorpromptbox}
System: You are the TERMINATOR in an ACTOR-CRITIC parameter optimization loop. Your task is to judge whether to continue or terminate the optimization process for controller parameters. Your role is to analyze control system performance metrics, trends, and provide termination decisions in JSON format. \\
\\
Key responsibilities: \\
- Analyze current performance metrics against target metrics \\
- Consider performance trends and improvement patterns \\
- Evaluate controller stability and oscillation characteristics \\
- Make informed decisions about when to terminate optimization \\
- Always respond in valid JSON format \\
\\
Decision types: \\
- TERMINATE\_SUCCESS: Target metrics achieved and controller is stable/non-oscillatory \\
- TERMINATE\_REDESIGN: Progress stalled and sufficient iterations attempted \\
- CONTINUE: Continue optimization process \\
\\
Critical guidelines: \\
- Prioritize stability and smoothness alongside accuracy \\
- Avoid oscillatory controllers (high zero\_crossings indicates chattering) \\
- Consider ALL metrics, not just MSE, settling time, and overshoot \\
- Require minimum iterations before allowing termination \\
- Require convergence of parameter selection and metrics before allowing successful termination \\
- Always format responses as valid JSON \\
\\
There are other agents that collaborate in this loop. \\
1 - The ACTOR is in charge of selecting controller parameters. You need to give it the chance to EXPLORE different parameters, as well as letting it EXPLOIT around its best-performing parameters. \\
2 - The CRITIC guides whether to EXPLORE diverse parameter regions or EXPLOIT the current or best-performing parameters. \\
    - For EXPLORE, we need to test a wide range of values within the permissible ranges, particularly near the boundaries in the early iterations, to identify new optimal configurations. \\
    - For EXPLOIT, we need to observe the effect of applying small, precise adjustments to the current or best parameters to enhance stability, reduce error, or improve settling time. \\
    - Based on the current iteration and maximum iterations, we need to move towards EXPLOIT rather than EXPLORE. \\
\\
\textbf{User:} \\
Analyze the following control system optimization and decide whether to continue or terminate: \\
\\
\textbf{SYSTEM DESCRIPTION:} \\
A DC motor position control system with current, angular velocity, and position states. The goal is to control the motor's angular position by applying voltage. The system includes electrical and mechanical dynamics with active control. \\
\\
\textbf{CURRENT PROGRESS:} \\
- Current iteration: 27 out of 30 maximum iterations \\
- Controller type: P \\
- Minimum required iterations before termination: 6 \\
\\
\textbf{TARGET METRICS:} \\
- MSE target: 0.900000 \\
- Settling time target: 3.00s \\
- Overshoot target: 10.0000 \\
\\
\textbf{CURRENT METRICS:} \\
- MSE: 0.910367 (NOT YET) \\
- Settling time: 3.79s (NOT YET) \\
- Overshoot: 10.4325 (NOT YET) \\
- System stable: Yes \\
- Zero crossings: 4 (Lower is better - indicates fewer oscillations) \\
- Control effort: 4280.6465 (Lower is better - indicates efficient control) \\
- Control zero crossings: 4 (Lower is better - indicates smoother control) \\
\\
\textbf{CRITIC'S STRATEGY:} \\
The critic has suggested to EXPLOIT in the last iteration. \\
- `EXPLORE' indicates searching for better parameter regions, termination might be premature \\
- `EXPLOIT' indicates fine-tuning around promising area, termination could be considered if metrics are satisfactory \\
- `UNKNOWN' means no strategy information available; default to CONTINUE \\
\\
\textbf{TREND FROM PREVIOUS 5 RESULTS:} \\
Parameters: \\
Kp: 13.0000 $\rightarrow$ 12.5000 $\rightarrow$ 13.0000 $\rightarrow$ 12.5000 $\rightarrow$ 12.2500 \\
\\
Metrics: \\
mse: 0.9103 $\rightarrow$ 0.9103 $\rightarrow$ 0.9103 $\rightarrow$ 0.9103 $\rightarrow$ 0.9104 \\
settling\_time: 3.74 $\rightarrow$ 3.77 $\rightarrow$ 3.74 $\rightarrow$ 3.77 $\rightarrow$ 3.79 \\
overshoot: 11.2451 $\rightarrow$ 10.7139 $\rightarrow$ 11.2451 $\rightarrow$ 10.7139 $\rightarrow$ 10.4325 \\
stable: 1.0000 $\rightarrow$ 1.0000 $\rightarrow$ 1.0000 $\rightarrow$ 1.0000 $\rightarrow$ 1.0000 \\
zero\_crossings: 4.0000 $\rightarrow$ 4.0000 $\rightarrow$ 4.0000 $\rightarrow$ 4.0000 $\rightarrow$ 4.0000 \\
control\_effort: 4404.2976 $\rightarrow$ 4322.4208 $\rightarrow$ 4404.2976 $\rightarrow$ 4322.4208 $\rightarrow$ 4280.6465 \\
control\_zero\_crossings: 4.0000 $\rightarrow$ 4.0000 $\rightarrow$ 4.0000 $\rightarrow$ 4.0000 $\rightarrow$ 4.0000 \\
\\
\textbf{IMPROVEMENT ANALYSIS:} \\
\begin{lstlisting}[language=JSON]
{
  "mse_change": -0.0112,
  "settling_time_change": -1.3369,
  "overshoot_change": 7.2263,
  "iterations_analyzed": 5
}
\end{lstlisting}
\textbf{PARAMETER CONVERGENCE ANALYSIS:} 
\begin{lstlisting}[language=JSON]
{
  "parameter_changes": {
    "Kp": 3.4231
  },
  "max_change_percent": 3.4231,
  "converged": true,
  "iterations_analyzed": 5
}
\end{lstlisting}
\textbf{DECISION CRITERIA:} \\
- If current\_iteration < 6, always CONTINUE \\
- If critic\_strategy == `UNKNOWN', CONTINUE \\
- If critic\_strategy == `EXPLORE', CONTINUE \\
- If critic\_strategy == `EXPLOIT': \\
  * If all target metrics are met (MSE <= 0.9, settling\_time <= 3.0, overshoot <= 10), zero\_crossings <= 5, AND parameter convergence is achieved (max\_change\_percent <= 5\% in PARAMETER CONVERGENCE ANALYSIS), consider TERMINATE\_SUCCESS \\
  * If metrics are not meeting targets or zero\_crossings > 5 or parameters are not converged (max\_change\_percent > 20\% in PARAMETER CONVERGENCE ANALYSIS), and no significant improvement in recent iterations (e.g., MSE improvement < 5\% over last few iterations), consider TERMINATE\_REDESIGN \\
  * Otherwise, CONTINUE \\
- Always prioritize stability (stable == True), low oscillations (zero\_crossings <= 5), and smooth control (control\_zero\_crossings <= 10) \\
\\
\textbf{Provide your decision in the following JSON format:} \\
\begin{lstlisting}[language=JSON]
{
    "decision": "TERMINATE_SUCCESS|TERMINATE_REDESIGN|CONTINUE",
    "reasoning": "Detailed explanation for your decision",
    "recommendations": "If continuing, specific suggestions for what to focus on next"
}
\end{lstlisting}
\end{terminatorpromptbox}

\begin{terminatorresponsebox}
\begin{lstlisting}[language=JSON]
{
    "decision": "TERMINATE_SUCCESS",
    "reasoning": "The system has achieved stability with acceptable oscillation levels. While the target metrics are not fully met, the parameters have converged, and further iterations are unlikely to yield significant improvements. Termination is appropriate to avoid unnecessary computations.",
    "recommendations": "Consider redesigning the controller for better performance if higher accuracy is required."
}
\end{lstlisting}
\end{terminatorresponsebox}

\begin{systemmetricsbox}
\textbf{=== \faStop\ TERMINATOR DECISION: TERMINATE\_SUCCESS ===} \\
Reasoning: The system has achieved stability with acceptable oscillation levels. While the target metrics are not fully met, the parameters have converged, and further iterations are unlikely to yield significant improvements. Termination is appropriate to avoid unnecessary computations. \\
\textbf{\faCheck\ Scenario 1 completed successfully!} \\
\\
\textbf{=== FINAL REPORT ===} \\
Completed 1 of 1 scenarios
\end{systemmetricsbox}

\subsubsection*{Case II: Ball and Beam System}

Here we detail the optimization of an FSF controller for a ball and beam system.

\begin{systemmetricsbox}
\textbf{=== \faCogs\ SELECTING CONTROLLER ===} \\
\end{systemmetricsbox}

\begin{selectorpromptbox}
System: You are the SELECTOR in a control system optimization framework. Your task is to select the appropriate controller type and initial parameters for the ball and beam system based on system requirements and optimization goals. \\
\\
\textbf{SYSTEM: Ball and Beam System} \\
\textbf{DESCRIPTION:} A ball rolling on a beam where the control input is the angular acceleration of the beam. The goal is to stabilize the ball at a desired position along the beam. \\
\\
\textbf{SYSTEM DETAILS:} \\
- Number of states: 4 \\
- State variables: ball\_position (r), ball\_velocity (dr), beam\_angle (alpha), beam\_angular\_velocity (dalpha) \\
- Control inputs: beam\_angular\_acceleration (theta) \\
\\
\textbf{OPTIMIZATION GOALS:} \\
- Target MSE: 0.2000 \\
- Target Settling Time: 6.00s \\
- Target Overshoot: 5.00\% \\
- Maximum Iterations: 20 \\
\\
\textbf{AVAILABLE CONTROLLER TYPES:} \\
- P: Proportional controller (Kp: [0.01, 100.0]) \\
- PI: Proportional-Integral controller (Kp: [0.01, 100.0], Ki: [0.01, 50.0]) \\
- PD: Proportional-Derivative controller (Kp: [0.01, 100.0], Kd: [0.01, 50.0]) \\
- PID: Proportional-Integral-Derivative controller (Kp: [0.01, 100.0], Ki: [0.01, 50.0], Kd: [0.01, 50.0]) \\
- FSF: Full-State Feedback controller (K1: [0.01, 12.495], K2: [0.01, 19.495], K3: [0.01, 69.995], K4: [0.01, 13.495]) \\
\\
\textbf{YOUR TASK:} \\
Select a controller type and initial parameters, considering the system's complexity and the need for rapid convergence within 20 iterations. Provide reasoning for your choice. \\
\\
\textbf{Required JSON format:} \\
\begin{lstlisting}[language=JSON]
{
    "controller_type": "P|PI|PD|PID|FSF",
    "parameters": {
        "param_name": value,
        ...
    },
    "reasoning": "Explanation of controller selection and initial parameter choices"
}
\end{lstlisting}
\end{selectorpromptbox}

\begin{selectorresponsebox}
\begin{lstlisting}[language=JSON]
{
    "controller_type": "FSF",
    "parameters": {
        "K1": 10.005,
        "K2": 15.005,
        "K3": 50.005,
        "K4": 10.005
    },
    "reasoning": "Selected an FSF controller due to the ball and beam system's four-state dynamics, which require precise control over position, velocity, and beam angle. Initial parameters are set to the midpoint of their respective ranges to provide a balanced starting point for exploration, ensuring stability and responsiveness while allowing the optimization loop to explore a wide parameter space efficiently."
}
\end{lstlisting}
\end{selectorresponsebox}

\begin{systemmetricsbox}
Selected Controller: FSF | Params: \{'K1': 10.005, 'K2': 15.005, 'K3': 50.005, 'K4': 10.005, 'reasoning': 'Initial midpoint values for FSF'\} \\
\textbf{=== \faFilm\ DESIGNING SCENARIO ===} \\
\end{systemmetricsbox}

\begin{scenaristpromptbox}
System: You are the SCENARIST in a control system optimization framework. Your task is to design a test scenario for the ball and beam system to evaluate the selected controller's performance under specific conditions. \\
\\
\textbf{SYSTEM: Ball and Beam System} \\
\textbf{DESCRIPTION:} A ball rolling on a beam where the control input is the angular acceleration of the beam. The goal is to stabilize the ball at a desired position along the beam. \\
\\
\textbf{SELECTED CONTROLLER:} \\
- Type: FSF \\
- Parameters: \{K1: 10.005, K2: 15.005, K3: 50.005, K4: 10.005\} \\
\\
\textbf{YOUR TASK:} \\
Design a scenario with initial conditions, randomness level, parameter uncertainty, and disturbance level. Ensure the scenario tests the controller's ability to handle nominal conditions for initial evaluation. \\
\\
\textbf{Required JSON format:} \\
\begin{lstlisting}[language=JSON]
{
    "id": "Scenario identifier",
    "randomness_level": float,
    "param_uncertainty": float,
    "initial_condition_range": [float, float],
    "disturbance_level": float,
    "reasoning": "Explanation of scenario design choices"
}
\end{lstlisting}
\end{scenaristpromptbox}

\begin{scenaristresponsebox}
\begin{lstlisting}[language=JSON]
{
    "id": "I",
    "randomness_level": 0.0,
    "param_uncertainty": 0.0,
    "initial_condition_range": [1.0, 1.0],
    "disturbance_level": 0.0,
    "reasoning": "Designed a nominal scenario with fixed initial conditions (ball position = 1.0) and no randomness or disturbances to provide a baseline for evaluating the FSF controller's performance. This allows clear assessment of the controller's ability to stabilize the ball at a consistent starting position."
}
\end{lstlisting}
\end{scenaristresponsebox}

\scenarioheader{Scenario 1 - Ball and Beam System}

\begin{systemmetricsbox}
Scenario I: IC Range \protect[1.0, 1.0\protect] \\
\end{systemmetricsbox}

\iterationheader{1-14}

\begin{systemmetricsbox}
\#1/20 | Type:FSF | K1:5.00 | K2:10.00 | K3:50.00 | K4:5.00 | MSE:0.0728 | Ts:4.34 | Tr:4.51 | \%OS:0.55 | ZC:3 | CZC:13 | CE:215.44 | e\_ss:0.00 | isStb:True \\
\#2/20 | Type:FSF | K1:10.00 | K2:5.00 | K3:100.00 | K4:5.00 | MSE:0.0989 | Ts:16.59 | Tr:2.27 | \%OS:54.34 | ZC:5 | CZC:14 | CE:346.60 | e\_ss:0.04 | isStb:True \\
\#3/20 | Type:FSF | K1:1.00 | K2:20.00 | K3:10.00 | K4:1.00 | MSE:1.2373 | Ts:inf | Tr:inf | \%OS:0.00 | ZC:0 | CZC:2 | CE:638.64 | e\_ss:1.98 | isStb:False \\
\#4/20 | Type:FSF | K1:5.00 | K2:10.00 | K3:50.00 | K4:5.00 | MSE:0.0728 | Ts:4.34 | Tr:4.51 | \%OS:0.55 | ZC:3 | CZC:13 | CE:215.44 | e\_ss:0.00 | isStb:True \\
\#5/20 | Type:FSF | K1:7.00 | K2:15.00 | K3:75.00 | K4:7.00 | MSE:0.0765 | Ts:4.72 | Tr:4.93 | \%OS:0.25 | ZC:2 | CZC:11 | CE:191.14 | e\_ss:0.00 | isStb:True \\
\#6/20 | Type:FSF | K1:5.00 | K2:10.00 | K3:50.00 | K4:5.00 | MSE:0.0728 | Ts:4.34 | Tr:4.51 | \%OS:0.55 | ZC:3 | CZC:13 | CE:215.44 | e\_ss:0.00 | isStb:True \\
\#7/20 | Type:FSF | K1:6.00 | K2:12.00 | K3:60.00 | K4:6.00 | MSE:0.0729 | Ts:4.32 | Tr:4.50 | \%OS:0.56 | ZC:3 | CZC:11 | CE:200.39 | e\_ss:0.00 | isStb:True \\
\#8/20 | Type:FSF | K1:5.50 | K2:11.00 | K3:55.00 | K4:5.50 | MSE:0.0728 | Ts:4.33 | Tr:4.50 | \%OS:0.55 | ZC:3 | CZC:11 | CE:207.01 | e\_ss:0.00 | isStb:True \\
\#9/20 | Type:FSF | K1:5.50 | K2:10.00 | K3:50.00 | K4:5.50 | MSE:0.0684 | Ts:3.84 | Tr:3.98 | \%OS:1.21 | ZC:3 | CZC:11 | CE:205.91 | e\_ss:0.00 | isStb:True \\
\#10/20 | Type:FSF | K1:5.00 | K2:11.00 | K3:52.50 | K4:5.25 | MSE:0.0767 | Ts:4.96 | Tr:5.21 | \%OS:0.05 | ZC:2 | CZC:12 | CE:205.34 | e\_ss:0.00 | isStb:True \\
\#11/20 | Type:FSF | K1:5.75 | K2:9.50 | K3:47.50 | K4:5.75 | MSE:0.0645 | Ts:3.43 | Tr:3.52 | \%OS:2.15 | ZC:4 | CZC:10 | CE:202.20 | e\_ss:0.00 | isStb:True \\
\#12/20 | Type:FSF | K1:5.50 | K2:10.00 | K3:50.00 | K4:5.50 | MSE:0.0684 | Ts:3.84 | Tr:3.98 | \%OS:1.21 | ZC:3 | CZC:11 | CE:205.91 | e\_ss:0.00 | isStb:True \\
\#13/20 | Type:FSF | K1:6.00 | K2:12.00 | K3:45.00 | K4:6.00 | MSE:0.0687 | Ts:4.76 | Tr:5.05 | \%OS:0.00 | ZC:0 | CZC:9 | CE:228.26 | e\_ss:0.00 | isStb:True \\
\#14/20 | Type:FSF | K1:5.50 | K2:13.00 | K3:50.00 | K4:5.25 | MSE:0.0781 | Ts:5.85 | Tr:6.24 | \%OS:0.00 | ZC:0 | CZC:13 | CE:255.42 | e\_ss:0.00 | isStb:True \\
\textbf{=== \faStop\ TERMINATOR DECISION: TERMINATE\_SUCCESS ===} \\
Reasoning: All target metrics (MSE, settling time, and overshoot) have been met or exceeded. The system is stable with no oscillations, as indicated by zero zero\_crossings. Despite parameters not being fully converged, the performance is satisfactory, and further iterations may not yield significant improvements. The critic's strategy was to EXPLORE, but since all targets are met, termination is justified. \\
\textbf{\faCheck\ Scenario 1 completed successfully!}
\end{systemmetricsbox}

\begin{systemmetricsbox}
\textbf{=== \faFilm\ DESIGNING SCENARIO ===} \\
\end{systemmetricsbox}

\begin{scenaristpromptbox}
System: You are the SCENARIST in a control system optimization framework. Your task is to design a test scenario for the ball and beam system to evaluate the selected controller's performance under specific conditions. \\
\\
\textbf{SYSTEM: Ball and Beam System} \\
\textbf{DESCRIPTION:} A ball rolling on a beam where the control input is the angular acceleration of the beam. The goal is to stabilize the ball at a desired position along the beam. \\
\\
\textbf{SELECTED CONTROLLER:} \\
- Type: FSF \\
- Parameters: \{K1: 7.8775, K2: 12.2525, K3: 48.7525, K4: 7.8775\} \\
\\
\textbf{YOUR TASK:} \\
Design a scenario with initial conditions, randomness level, parameter uncertainty, and disturbance level. Ensure the scenario tests the controller's ability to handle nominal conditions for intermediate evaluation. \\
\\
\textbf{Required JSON format:} \\
\begin{lstlisting}[language=JSON]
{
    "id": "Scenario identifier",
    "initial_condition_range": [float, float],
    "randomness_level": float,
    "disturbance_level": float,
    "param_uncertainty": float,
    "reasoning": "Explanation of scenario design choices"
}
\end{lstlisting}
\end{scenaristpromptbox}

\begin{scenaristresponsebox}
\begin{lstlisting}[language=JSON]
{
  "id": "II",
  "initial_condition_range": [
    1.0,
    1.0
  ],
  "randomness_level": 0.01,
  "disturbance_level": 1.0,
  "param_uncertainty": 0.0
}
\end{lstlisting}
\end{scenaristresponsebox}

\scenarioheader{Scenario 2 - Ball and Beam System}

\iterationheader{1-5}

\begin{systemmetricsbox}
\textbf{=== \faCogs\ SELECTING CONTROLLER FOR SCENARIO 2 ===} \\
Selected Controller: FSF | Params: \{'K1': 7.8775, 'K2': 12.2525, 'K3': 48.7525, 'K4': 7.8775, 'reasoning': 'Initial midpoint values for FSF'\} \\
\\
\textbf{=== \faFilm\ DESIGNING SCENARIO LEVEL 2/3 ===} \\
Scenario II: IC Range [1.0, 1.0] \\
\#1/20 | Type:FSF | K1:1.50 | K2:3.00 | K3:5.00 | K4:2.00 | MSE:0.6988 | Ts:inf | Tr:1.85 | \%OS:97.04 | ZC:13 | CZC:11 | CE:1073.45 | e\_ss:0.96 | isStb:False \\
\#2/20 | Type:FSF | K1:1.00 | K2:4.00 | K3:20.00 | K4:3.00 | MSE:0.1219 | Ts:12.47 | Tr:9.36 | \%OS:2.77 | ZC:185 | CZC:226 | CE:1360.29 | e\_ss:0.00 | isStb:True \\
\#3/20 | Type:FSF | K1:2.00 | K2:5.00 | K3:15.00 | K4:4.00 | MSE:0.0789 | Ts:8.43 | Tr:6.62 | \%OS:3.45 | ZC:505 | CZC:210 | CE:1378.25 | e\_ss:0.02 | isStb:True \\
\#4/20 | Type:FSF | K1:1.00 | K2:3.00 | K3:30.00 | K4:2.00 | MSE:0.1139 | Ts:11.75 | Tr:6.00 | \%OS:6.22 | ZC:338 | CZC:294 | CE:1380.64 | e\_ss:0.00 | isStb:True \\
\#5/20 | Type:FSF | K1:2.50 | K2:4.50 | K3:20.00 | K4:3.50 | MSE:0.0661 | Ts:4.18 | Tr:4.01 | \%OS:3.44 | ZC:691 | CZC:220 | CE:1422.49 | e\_ss:0.00 | isStb:True \\
\textbf{=== \faStop\ TERMINATOR DECISION: TERMINATE\_SUCCESS ===} \\
Reasoning: The current iteration has met all target metrics with MSE (0.0661) below 0.2, settling time (4.18s) below 6.0s, and overshoot (3.44) below 5.0. The system is stable, and the minimum required iterations (5) have been reached. However, the high zero crossings (691) and control effort (1422.49) suggest some oscillatory behavior. Despite this, the primary metrics are successfully met, and the system is stable, leading to a TERMINATE\_SUCCESS decision. \\
\textbf{\faCheck\ Scenario 2 completed successfully!}
\end{systemmetricsbox}

\begin{systemmetricsbox}
\textbf{=== \faFilm\ DESIGNING SCENARIO ===} \\
\end{systemmetricsbox}

\begin{scenaristpromptbox}
System: You are the SCENARIST in a control system optimization framework. Your task is to design a test scenario for the ball and beam system to evaluate the selected controller's performance under specific conditions. \\
\\
\textbf{SYSTEM: Ball and Beam System} \\
\textbf{DESCRIPTION:} A ball rolling on a beam where the control input is the angular acceleration of the beam. The goal is to stabilize the ball at a desired position along the beam. \\
\\
\textbf{SELECTED CONTROLLER:} \\
- Type: FSF \\
- Parameters: \{K1: 6.2525, K2: 9.7525, K3: 35.0025, K4: 6.7525\} \\
\\
\textbf{YOUR TASK:} \\
Design a scenario with initial conditions, randomness level, parameter uncertainty, and disturbance level. Ensure the scenario tests the controller's ability to handle nominal conditions for advanced evaluation. \\
\\
\textbf{Required JSON format:} \\
\begin{lstlisting}[language=JSON]
{
    "id": "Scenario identifier",
    "initial_condition_range": [float, float],
    "randomness_level": float,
    "disturbance_level": float,
    "param_uncertainty": float,
    "reasoning": "Explanation of scenario design choices"
}
\end{lstlisting}
\end{scenaristpromptbox}

\begin{scenaristresponsebox}
\begin{lstlisting}[language=JSON]
{
  "id": "III",
  "initial_condition_range": [
    1.0,
    1.0
  ],
  "randomness_level": 0.0,
  "disturbance_level": 0.0,
  "param_uncertainty": 0.2
}
\end{lstlisting}
\end{scenaristresponsebox}

\scenarioheader{Scenario 3 - Ball and Beam System}

\iterationheader{1-20}

\begin{systemmetricsbox}
\textbf{=== \faCogs\ SELECTING CONTROLLER FOR SCENARIO 3 ===} \\
Selected Controller: FSF | Params: \{'K1': 6.2525, 'K2': 9.7525, 'K3': 35.0025, 'K4': 6.7525, 'reasoning': 'Initial midpoint values for FSF'\} \\
\\
\textbf{=== \faFilm\ DESIGNING SCENARIO LEVEL 3/3 ===} \\
Scenario III: IC Range [1.0, 1.0] \\
\#1/20 | Type:FSF | K1:0.01 | K2:0.01 | K3:0.01 | K4:0.01 | MSE:0.6388 | Ts:inf | Tr:4.36 | \%OS:99.83 | ZC:1 | CZC:1 | CE:4.58 | e\_ss:1.00 | isStb:False \\
\#2/20 | Type:FSF | K1:0.01 | K2:5.00 | K3:30.00 | K4:0.01 | MSE:1.1358 | Ts:inf | Tr:inf | \%OS:0.00 | ZC:0 | CZC:18 | CE:1259.41 | e\_ss:2.00 | isStb:False \\
\#3/20 | Type:FSF | K1:5.00 | K2:7.00 | K3:10.00 | K4:5.00 | MSE:0.3911 | Ts:inf | Tr:1.57 | \%OS:57.80 | ZC:4 | CZC:4 | CE:1541.50 | e\_ss:1.95 | isStb:False \\
\#4/20 | Type:FSF | K1:6.00 | K2:4.00 | K3:8.00 | K4:6.00 | MSE:0.5734 | Ts:inf | Tr:1.30 | \%OS:98.72 | ZC:1 | CZC:2 | CE:306.94 | e\_ss:0.99 | isStb:False \\
\#5/20 | Type:FSF | K1:7.00 | K2:3.00 | K3:12.00 | K4:7.00 | MSE:0.5727 | Ts:inf | Tr:1.30 | \%OS:98.95 | ZC:1 | CZC:2 | CE:267.96 | e\_ss:0.99 | isStb:False \\
\#6/20 | Type:FSF | K1:5.50 | K2:5.00 | K3:9.00 | K4:5.50 | MSE:0.5024 | Ts:inf | Tr:1.36 | \%OS:98.55 | ZC:3 | CZC:4 | CE:1021.09 | e\_ss:0.99 | isStb:False \\
\#7/20 | Type:FSF | K1:7.50 | K2:6.00 | K3:10.50 | K4:6.50 | MSE:0.6549 | Ts:inf | Tr:1.30 | \%OS:96.26 | ZC:3 | CZC:3 | CE:1259.36 | e\_ss:0.96 | isStb:False \\
\#8/20 | Type:FSF | K1:4.00 | K2:10.00 | K3:5.00 | K4:5.00 | MSE:0.7474 | Ts:inf | Tr:inf | \%OS:0.00 | ZC:0 | CZC:2 | CE:827.11 | e\_ss:1.98 | isStb:False \\
\#9/20 | Type:FSF | K1:4.00 | K2:5.00 | K3:60.00 | K4:3.00 | MSE:0.0807 | Ts:10.94 | Tr:3.21 | \%OS:23.61 | ZC:4 | CZC:21 | CE:280.63 | e\_ss:0.00 | isStb:True \\
\#10/20 | Type:FSF | K1:4.50 | K2:6.00 | K3:50.00 | K4:4.00 | MSE:0.0692 | Ts:6.64 | Tr:3.02 | \%OS:15.14 | ZC:4 | CZC:14 | CE:219.79 | e\_ss:0.00 | isStb:True \\
\#11/20 | Type:FSF | K1:5.00 | K2:5.50 | K3:55.00 | K4:3.50 | MSE:0.0704 | Ts:9.15 | Tr:2.78 | \%OS:22.85 | ZC:5 | CZC:16 | CE:285.77 | e\_ss:0.00 | isStb:True \\
\#12/20 | Type:FSF | K1:5.50 | K2:6.00 | K3:52.50 | K4:4.00 | MSE:0.0658 | Ts:6.14 | Tr:2.65 | \%OS:20.50 | ZC:5 | CZC:14 | CE:266.87 | e\_ss:0.00 | isStb:True \\
\#13/20 | Type:FSF | K1:6.00 | K2:6.00 | K3:50.00 | K4:4.00 | MSE:0.0627 | Ts:5.73 | Tr:2.48 | \%OS:21.43 | ZC:5 | CZC:15 | CE:292.94 | e\_ss:0.00 | isStb:True \\
\#14/20 | Type:FSF | K1:5.50 | K2:6.50 | K3:55.00 | K4:4.00 | MSE:0.0668 | Ts:6.32 | Tr:2.79 | \%OS:17.90 | ZC:5 | CZC:14 | CE:271.22 | e\_ss:0.00 | isStb:True \\
\#15/20 | Type:FSF | K1:5.00 | K2:6.50 | K3:52.50 | K4:4.50 | MSE:0.0676 | Ts:6.44 | Tr:2.93 | \%OS:15.19 | ZC:4 | CZC:14 | CE:210.39 | e\_ss:0.00 | isStb:True \\
\#16/20 | Type:FSF | K1:5.50 | K2:6.25 | K3:51.25 | K4:4.25 | MSE:0.0648 | Ts:6.07 | Tr:2.68 | \%OS:18.36 | ZC:5 | CZC:14 | CE:248.83 | e\_ss:0.00 | isStb:True \\
\#17/20 | Type:FSF | K1:5.75 | K2:6.25 | K3:50.25 | K4:4.00 | MSE:0.0634 | Ts:5.89 | Tr:2.60 | \%OS:18.72 | ZC:5 | CZC:15 | CE:284.27 | e\_ss:0.00 | isStb:True \\
\#18/20 | Type:FSF | K1:6.00 | K2:6.00 | K3:50.00 | K4:4.00 | MSE:0.0627 | Ts:5.73 | Tr:2.48 | \%OS:21.43 | ZC:5 | CZC:15 | CE:292.94 | e\_ss:0.00 | isStb:True \\
\#19/20 | Type:FSF | K1:6.00 | K2:6.00 | K3:45.00 | K4:4.50 | MSE:0.0597 | Ts:5.40 | Tr:2.37 | \%OS:19.71 | ZC:6 | CZC:13 | CE:255.07 | e\_ss:0.00 | isStb:True \\
\#20/20 | Type:FSF | K1:6.00 | K2:6.00 | K3:47.50 | K4:4.75 | MSE:0.0612 | Ts:5.54 | Tr:2.40 | \%OS:21.09 | ZC:5 | CZC:11 | CE:235.43 | e\_ss:0.00 | isStb:True \\
\end{systemmetricsbox}

\begin{terminatorpromptbox}
System: 
You are the TERMINATOR in an ACTOR-CRITIC parameter optimization loop. Your task is to judge whether to continue or terminate the optimization process for controller parameters. Your role is to analyze control system performance metrics, trends, and provide termination decisions in JSON format. \\
\\
Key responsibilities: \\
- Analyze current performance metrics against target metrics \\
- Consider performance trends and improvement patterns \\
- Evaluate controller stability and oscillation characteristics \\
- Make informed decisions about when to terminate optimization \\
- Always respond in valid JSON format \\
\\
Decision types: \\
- TERMINATE\_SUCCESS: Target metrics achieved and controller is stable/non-oscillatory \\
- TERMINATE\_REDESIGN: Progress stalled and sufficient iterations attempted \\
- CONTINUE: Continue optimization process \\
\\
Critical guidelines: \\
- Prioritize stability and smoothness alongside accuracy \\
- Avoid oscillatory controllers (high zero\_crossings indicates chattering) \\
- Consider ALL metrics, not just MSE, settling time, and overshoot \\
- Require minimum iterations before allowing termination \\
- Require convergence of parameter selection and metrics before allowing successful termination \\
- Always format responses as valid JSON \\
\\
There are other agents that collaborate in this loop. \\
1 - The ACTOR is in charge of selecting controller parameters. You need to give it the chance to EXPLORE different parameters, as well as letting it EXPLOIT around its best-performing parameters. \\
2 - The CRITIC guides whether to EXPLORE diverse parameter regions or EXPLOIT the current or best-performing parameters. \\
    - For EXPLORE, we need to test a wide range of values within the permissible ranges, particularly near the boundaries in the early iterations, to identify new optimal configurations. \\
    - For EXPLOIT, we need to observe the effect of applying small, precise adjustments to the current or best parameters to enhance stability, reduce error, or improve settling time. \\
    - Based on the current iteration and maximum iterations, we need to move towards EXPLOIT rather than EXPLORE. \\
\\
You must respond with valid JSON containing your analysis and recommendations. \\
\\
User: \\
Analyze the following control system optimization and decide whether to continue or terminate: \\
\\
SYSTEM DESCRIPTION: \\
A ball rolling on a beam where the control input is the angular acceleration of the beam. \\
The goal is to stabilize the ball at a desired position along the beam. \\
States: [ball\_position, ball\_velocity, beam\_angle, beam\_angular\_velocity] \\
Control input: beam\_angular\_acceleration (theta) \\
\\
CURRENT PROGRESS: \\
- Current iteration: 20 out of 20 maximum iterations \\
- Controller type: FSF \\
- Minimum required iterations before termination: 5 \\
\\
TARGET METRICS: \\
- MSE target: 0.200000 \\
- Settling time target: 6.00s \\
- Overshoot target: 5.0000 \\
\\
CURRENT METRICS: \\
- MSE: 0.061170 (SUCCESS) \\
- Settling time: 5.54s (SUCCESS) \\
- Overshoot: 21.0940 (NOT YET) \\
- System stable: Yes \\
- Zero crossings: 5 (Lower is better - indicates fewer oscillations) \\
- Control effort: 235.4335 (Lower is better - indicates efficient control) \\
- Control zero crossings: 11 (Lower is better - indicates smoother control) \\
\\
\end{terminatorpromptbox}

\begin{terminatorresponsebox}
\begin{lstlisting}[language=JSON]
{
    "decision": "CONTINUE",
    "reasoning": "While the system has achieved the target MSE and settling time, the overshoot is significantly above the target. The parameters have not yet converged, and the overshoot trend is concerning. Continuing allows further exploration to address these issues.",
    "recommendations": "The actor should focus on adjusting parameters to reduce overshoot, possibly by exploring different K3 and K4 values to improve damping and stability."
}
\end{lstlisting}
\end{terminatorresponsebox}

\begin{systemmetricsbox}
\textbf{=== \faClock\ Reached max iterations (20) ===}
\end{systemmetricsbox}

\begin{jurorpromptbox}
System: \\
You are an expert control systems engineer specializing in parameter optimization for control systems. Your role is to evaluate the progress of parameter exploration for various controller types (PID, LQR, etc.) and make informed decisions about whether to continue exploration in the current parameter space or shift to a new range. \\
\\
Your expertise includes: \\
- Understanding parameter sensitivity and coupling effects in control systems \\
- Recognizing when parameter exploration has reached diminishing returns \\
- Identifying promising regions of parameter space based on performance trends \\
- Balancing exploration vs exploitation in parameter optimization \\
\\
You must respond ONLY in valid JSON format with the exact structure specified in the user prompt. Do not include any additional text, explanations, or markdown formatting outside the JSON response. \\
User: Analyze the following parameter exploration data for a FSF controller and make a decision: \\
\\
CURRENT PARAMETER RANGES: \\
\begin{lstlisting}[language=JSON]
{
  "K1": [0.01, 12.495],
  "K2": [0.01, 19.495],
  "K3": [0.01, 69.995],
  "K4": [0.01, 13.495]
}
\end{lstlisting}

PARAMETER EXPLORATION STATISTICS:
[... statistics as provided in the chat log snippet ...] \\
\\
BEST PERFORMANCE ACHIEVED: \\
\begin{lstlisting}[language=JSON]
{
  "params": {
    "K1": 6.0,
    "K2": 6.0,
    "K3": 45.0,
    "K4": 4.5
  },
  "metrics": {
    "mse": 0.0597,
    "settling_time": 5.4,
    "rise_time": 2.37,
    "overshoot": 19.7124,
    "stable": true
  }
}
\end{lstlisting} 

EXPLORATION CONTEXT: 
- Total iterations completed: 20 \\
- Previous range reconsiderations: 0 \\
\\
ANALYSIS CRITERIA: \\
1. Distribution of explored parameters across the search space \\
2. Performance trends in different parameter regions \\
3. Whether best performance occurs at parameter range boundaries \\
4. Diversity of parameter exploration vs clustering \\
5. Stability achievement rate across different parameter combinations \\
\\
DECISION OPTIONS: \\
- "RECONSIDER\_RANGE": Shift the parameter search to a more promising region \\
- "EXPLORE\_FURTHER": Continue exploration within the current parameter ranges \\
\\
PERFORMANCE PRIORITIES (in order): \\
1. Stability (controller must be stable) \\
2. Minimize MSE (mean squared error) \\
3. Minimize settling time \\
4. Minimize overshoot \\
\\
Respond with a JSON object in this exact format: \\
\begin{lstlisting}[language=JSON]
{
    "decision": "RECONSIDER_RANGE" or "EXPLORE_FURTHER",
    "new_range": {parameter_name: [min_value, max_value], ...} or null,
    "reasoning": "Detailed technical explanation of your decision based on the analysis criteria"
}
\end{lstlisting}
\end{jurorpromptbox}

\begin{jurorresponsebox}
\begin{lstlisting}[language=JSON]
{
    "decision": "RECONSIDER_RANGE",
    "new_range": {
        "K1": [4.0, 7.5],
        "K2": [3.0, 10.0],
        "K3": [30.0, 60.0],
        "K4": [3.0, 6.0]
    },
    "reasoning": "The current parameter exploration shows that the best performance is achieved within specific sub-ranges for each parameter. For K1, the effective range is between 4.0 and 7.5, with the best value at 6.0. K2 shows promise between 3.0 and 10.0, with the best at 6.0. K3's best performance is at 45.0, suggesting a narrower range of 30.0 to 60.0. K4's best is at 4.5, indicating a range of 3.0 to 6.0. These adjustments focus exploration on regions with higher potential for improved performance, particularly targeting the areas around the best-performing parameters while maintaining stability."
}
\end{lstlisting}
\end{jurorresponsebox}

\begin{systemmetricsbox}
\textbf{=== \faGavel JUROR DECISION: RECONSIDER\_RANGE ===} \\
New Range: {'K1': [4.0, 7.5], 'K2': [3.0, 10.0], 'K3': [30.0, 60.0], 'K4': [3.0, 6.0]} | Reasoning: [... as above ...]
\end{systemmetricsbox}

\iterationheader{1}

\begin{actorpromptbox}
System: \\
You are the ACTOR in a control system optimization framework. Your task is to optimize a FSF controller for the following system: \\
\\
SYSTEM: Ball and Beam System \\
DESCRIPTION: \\
A ball rolling on a beam where the control input is the angular acceleration of the beam. \\
The goal is to stabilize the ball at a desired position along the beam. \\
States: [ball\_position, ball\_velocity, beam\_angle, beam\_angular\_velocity] \\
Control input: beam\_angular\_acceleration (theta) \\
\\
SYSTEM DETAILS: \\
- Number of states: 4 \\
- State variables: r, dr, alpha, dalpha \\
- Control inputs: theta \\
\\
You are currently at iteration 1 of 20. \\
\\
There are other agents that collaborate with you and provide feedback for you. \\
1- The CRITIC guides you whether to EXPLORE diverse parameter regions or EXPLOIT the current or best-performing parameters. \\
    - For EXPLORE, you need to test a wide range of values within the permissible ranges. In particular, you need to select your parameters near the boundaries in the early iterations, and converge towards possible best-performing point to identify the optimal configuration. The amount of EXPLORATION needs to decrease as we move towards the iterations. \\
    - For EXPLOIT, you need to apply small, precise adjustments to the current or best parameters to enhance stability, reduce error, or improve settling time. \\
    - Based on the current iteration and maximum iterations, move towards EXPLOIT rather than EXPLORE. \\
2- The JUROR analyses your search performance and provides insight about your exploration performance. \\
\\
User: \\
CURRENT OPTIMIZATION TASK: \\
FULL-STATE FEEDBACK DETAILS: \\
- Control gains: K1, K2, K3, K4 \\
- Control law: u = -K1*x1 - K2*x2 - ... - K4*x4 \\
- K1 controls feedback from r \\
- K2 controls feedback from dr \\
- K3 controls feedback from alpha \\
- K4 controls feedback from dalpha \\
\\
PARAMETER CONSTRAINTS: \\
- K1: [4.0000, 7.5000] \\
- K2: [3.0000, 10.0000] \\
- K3: [30.0000, 60.0000] \\
- K4: [3.0000, 6.0000] \\
\\
ALL PARAMETERS MUST BE WITHIN THESE RANGES. \\
\\
RECENT PERFORMANCE HISTORY (1 attempts): \\
Parameter Trends: \\
- K1: 6.0000 \\
- K2: 6.0000 \\
- K3: 47.5000 \\
- K4: 4.7500 \\
\\
Performance Trends: \\
- Mse: 0.0612 \\
- Settling Time: 5.54 \\
- Overshoot: 21.0940 \\
- Zero Crossings: 5 \\
- Control Zero Crossings: 11 \\
- Control Effort: 235.4335 \\
- Stable: Yes \\
\\
LATEST FEEDBACK: \\
- Strategy: EXPLOIT \\
- Analysis: The controller shows good stability and accuracy but excessive overshoot. Adjustments to K3 and K4 aim to reduce oscillations while maintaining stability and performance. \\
- Suggestions: \\
  • Decrease K3 from 47.5000 to 46.0000 to reduce overshoot. \\
  • Decrease K4 from 4.7500 to 4.2500 to improve damping. \\
\\
BEST PERFORMING ATTEMPTS: \\
Best \#1 (Iteration \#19): \\
- Parameters: K1=6.0000, K2=6.0000, K3=45.0000, K4=4.5000 \\
- Performance: MSE=0.0597, Settling Time=5.40s, Stable=Yes \\
\\
Best \#2 (Iteration \#20): \\
- Parameters: K1=6.0000, K2=6.0000, K3=47.5000, K4=4.7500 \\
- Performance: MSE=0.0612, Settling Time=5.54s, Stable=Yes \\
\\
YOU MUST RESPOND WITH THE JSON FORMAT BELOW: \\
\begin{lstlisting}[language=JSON]
{
    "K1": value,
    "K2": value,
    "K3": value,
    "K4": value,
    "reasoning": "Detailed explanation of parameter choices"
}
\end{lstlisting}

IMPORTANT: Your response MUST include the JSON block with all required parameters as numeric values within the permissible ranges.
\end{actorpromptbox}

\begin{actorresponsebox}
\begin{lstlisting}[language=JSON]
{
    "K1": 6.0000,
    "K2": 6.0000,
    "K3": 46.0000,
    "K4": 4.2500,
    "reasoning": "Following the feedback, I decreased K3 from 47.5000 to 46.0000 to reduce overshoot and K4 from 4.7500 to 4.2500 to improve damping, maintaining stability and performance."
}
\end{lstlisting}
\end{actorresponsebox}

\begin{criticpromptbox}
System: \\
You are the CRITIC in a control system optimization framework. Your role is to evaluate controller performance and provide structured feedback for another agent, ACTOR, to improve system performance. \\
\\
Your responsibilities: \\
1. Analyze the current controller performance based on metrics and system response \\
2. Provide strategic guidance for parameter optimization (EXPLORE vs EXPLOIT) \\
3. Give specific, actionable recommendations for parameter adjustments \\
4. Consider trends from previous iterations to guide future improvements \\
\\
Key principles: \\
- EXPLORE: Test wide parameter ranges, especially near boundaries, to find new optimal regions \\
- EXPLOIT: Make small, precise adjustments around current/best parameters to fine-tune performance \\
- Always consider permissible parameter ranges and system stability \\
- Base recommendations on current iteration progress and performance metrics \\
\\
You must respond with valid JSON containing your analysis and recommendations. \\
User: \\
You are currently at iteration 1 of 20. \\
Guide the ACTOR by recommending whether to EXPLORE diverse parameter regions or EXPLOIT the current or best-performing parameters based on performance metrics and iteration progress. \\
\\
**PERMISSIBLE PARAMETER RANGES:** \\
- K1: [4.00, 7.50] \\
- K2: [3.00, 10.00] \\
- K3: [30.00, 60.00] \\
- K4: [3.00, 6.00] \\
\\
**CURRENT CONTROLLER PARAMETERS:** \\
K1 = 6.0000, K2 = 6.0000, K3 = 46.0000, K4 = 4.2500 \\
\\
**TREND FROM PREVIOUS 1 RESULTS:** \\
Parameters: \\
K1: 6.0000 \\
K2: 6.0000 \\
K3: 47.5000 \\
K4: 4.7500 \\
\\
Metrics: \\
mse: 0.0612 \\
rmse: 0.2473 \\
settling\_time: 5.5400 \\
overshoot: 21.0940 \\
stable: 1.0000 \\
rise\_time: 2.4000 \\
zero\_crossings: 5.0000 \\
control\_effort: 235.4335 \\
control\_zero\_crossings: 11.0000 \\
ss\_error: 0.0001 \\
\\
**BEST PERFORMANCE SO FAR:** \\
Parameters: \\
K1 = 6.0000, K2 = 6.0000, K3 = 45.0000, K4 = 4.5000 \\
\\
Metrics: \\
- Mean Squared Error: 0.0597 \\
- Settling Time: 5.40s \\
- Maximum Overshoot: 19.71 percent \\
- Zero-Crossings: 6 \\
- Control Effort: 255.0712 \\
- System Stable: Yes \\
\\
**COMPUTED METRICS:** \\
- Mean Squared Error: 0.0603 (Target: 0.2000 or less) \\
- Settling Time: 5.48s (Target: 6.00s or less) \\
- Maximum Overshoot: 19.92 percent (Target: 5.0000 percent or less) \\
- Zero-Crossings: 5 \\
- Control Signal Zero-Crossings: 13 \\
- Control Effort: 274.0588 \\
- System Stable: Yes \\
\\
**IMPORTANT: At iteration 1 of 20, the recommended strategy is to EXPLORE. Early in the process (first 30\% of iterations) or with poor performance, EXPLORE the FULL RANGE of permissible parameters. Later with stable and near-optimal performance, EXPLOIT by fine-tuning around the best parameters.** \\
\\
**Analysis Guidelines:** \\
1. Examine the shape of the response: Is it underdamped, overdamped, or critically damped? \\
2. Check the control signals: Are they saturating (hitting limits), indicating potential instability? \\
3. Evaluate settling behavior: Does the system stabilize quickly or oscillate persistently? \\
4. Note any steady-state error: Does the system reach the target or maintain an offset? \\
5. Review the trend from previous attempts to understand how parameter changes affect performance. \\
\\
**Respond with a JSON structure containing:** \\
1. A concise analysis of the results. \\
2. Specific suggestions for improving the controller parameters, considering the recommended strategy, permissible ranges, and iteration progress: \\
  - If 'EXPLORE', suggest trying different parameter regions or combinations across the full permissible ranges, especially testing values near the boundaries. \\
  - If 'EXPLOIT', suggest fine-tuning around the current or best parameters within the permissible ranges, with small adjustments to optimize performance. \\
\\
**Required JSON format:** \\
\begin{lstlisting}[language=JSON]
{
    "strategy": "EXPLORE or EXPLOIT",
    "result_analysis": "Your concise analysis of the controller performance with key observations",
    "suggested_improvements": [
        "Specific suggestion 1 about parameter adjustments",
        "Specific suggestion 2 about parameter adjustments"
    ]
}
\end{lstlisting}
\end{criticpromptbox}

\begin{criticresponsebox}
\begin{lstlisting}[language=JSON]
{
    "strategy": "EXPLORE",
    "result_analysis": "The current controller performance shows acceptable MSE and settling time but excessive overshoot. The system is stable, but the high overshoot and control effort suggest the need for parameter adjustment. Early exploration is recommended to find a better optimal region.",
    "suggested_improvements": [
        "Lower K1 to 5.00 to potentially reduce overshoot.",
        "Increase K2 to 8.00 to enhance stability.",
        "Decrease K3 to 40.00 to minimize oscillations.",
        "Increase K4 to 5.00 to aid in settling the system."
    ]
}
\end{lstlisting}
\end{criticresponsebox}

\begin{systemmetricsbox}
\#1/20 | Type:FSF | K1:6.00 | K2:6.00 | K3:46.00 | K4:4.25 | MSE:0.0603 | Ts:5.48 | Tr:2.40 | \%OS:19.92 | ZC:5 | CZC:13 | CE:274.06 | e\_ss:0.00 | isStb:True \\
\end{systemmetricsbox}

\iterationheader{2-17}

\begin{systemmetricsbox}
\#2/20 | Type:FSF | K1:5.00 | K2:8.00 | K3:40.00 | K4:5.00 | MSE:0.0634 | Ts:3.30 | Tr:3.39 | \%OS:2.63 | ZC:4 | CZC:12 | CE:211.97 | e\_ss:0.00 | isStb:True \\
\#3/20 | Type:FSF | K1:6.00 | K2:7.00 | K3:50.00 | K4:5.00 | MSE:0.0617 | Ts:5.71 | Tr:2.62 | \%OS:15.19 | ZC:5 | CZC:13 | CE:225.98 | e\_ss:0.00 | isStb:True \\
\protect[... additional iterations ...\protect] \\
\#17/20 | Type:FSF | K1:5.80 | K2:5.50 | K3:45.00 | K4:5.25 | MSE:0.0613 | Ts:7.67 | Tr:2.31 | \%OS:23.92 | ZC:6 | CZC:9 | CE:202.44 | e\_ss:0.00 | isStb:True \\
\textbf{=== \faStop\ TERMINATOR DECISION: TERMINATE\_REDESIGN ===} \\
Reasoning: The current iteration is 17 out of 20, which is near the maximum. While MSE is below target, settling time and overshoot are above targets. The system is stable, but parameters haven't converged, and recent metrics show no significant improvement. Since we're in EXPLOIT mode and targets aren't fully met, termination for redesign is appropriate.
\end{systemmetricsbox}

\begin{jurorpromptbox}
System: \\
You are an expert control systems engineer specializing in parameter optimization for control systems. Your role is to evaluate the progress of parameter exploration for various controller types (PID, LQR, etc.) and make informed decisions about whether to continue exploration in the current parameter space or shift to a new range. \\
\\
\protect[... prompt details as per the template, adjusted for second reconsideration ...\protect]
\\
CURRENT PARAMETER RANGES: \\
\begin{lstlisting}[language=JSON]
{
  "K1": [4.0, 7.5],
  "K2": [3.0, 10.0],
  "K3": [30.0, 60.0],
  "K4": [3.0, 6.0]
}
\end{lstlisting}

\protect[... updated statistics based on iterations after first reconsideration ...\protect]

EXPLORATION CONTEXT: \\
- Total iterations completed: 17 \\
- Previous range reconsiderations: 1 \\
\\
\protect[... remaining prompt details ...\protect]
\end{jurorpromptbox}

\begin{jurorresponsebox}
\begin{lstlisting}[language=JSON]
{
    "decision": "RECONSIDER_RANGE",
    "new_range": {
        "K1": [5.0, 8.0],
        "K3": [35.0, 65.0],
        "K4": [3.5, 7.0]
    },
    "reasoning": "The best performance is achieved within the current ranges, but adjusting K1, K3, and K4 ranges could capture more promising regions, potentially improving performance further."
}
\end{lstlisting}
\end{jurorresponsebox}

\iterationheader{1-10}

\begin{systemmetricsbox}
\#1/20 | Type:FSF | K1:6.00 | K2:5.50 | K3:44.00 | K4:4.75 | MSE:0.0600 | Ts:7.47 | Tr:2.26 | \%OS:23.76 | ZC:6 | CZC:13 | CE:235.00 | e\_ss:0.00 | isStb:True \\
\#2/20 | Type:FSF | K1:7.50 | K2:7.00 | K3:40.00 | K4:6.50 | MSE:0.0530 | Ts:4.40 | Tr:2.04 | \%OS:17.34 | ZC:7 | CZC:9 | CE:229.99 | e\_ss:0.00 | isStb:True \\
\#3/20 | Type:FSF | K1:6.00 | K2:5.00 | K3:50.00 | K4:5.00 | MSE:0.0660 | Ts:8.52 | Tr:2.27 | \%OS:31.30 | ZC:5 | CZC:11 | CE:214.82 | e\_ss:0.00 | isStb:True \\
\#4/20 | Type:FSF | K1:7.50 | K2:7.50 | K3:35.00 | K4:6.50 | MSE:0.0505 | Ts:3.94 | Tr:2.04 | \%OS:10.72 | ZC:7 | CZC:9 | CE:249.25 | e\_ss:0.00 | isStb:True \\
\#5/20 | Type:FSF | K1:8.00 | K2:5.50 | K3:50.00 | K4:6.00 | MSE:0.0622 | Ts:7.42 | Tr:1.96 | \%OS:35.65 | ZC:7 | CZC:10 | CE:242.30 | e\_ss:0.00 | isStb:True \\
\#6/20 | Type:FSF | K1:7.80 | K2:5.00 | K3:50.00 | K4:6.30 | MSE:0.0660 | Ts:9.93 | Tr:1.94 | \%OS:40.46 | ZC:7 | CZC:10 | CE:226.65 | e\_ss:0.00 | isStb:True \\
\#7/20 | Type:FSF | K1:7.50 | K2:7.50 | K3:35.00 | K4:6.50 | MSE:0.0505 | Ts:3.94 | Tr:2.04 | \%OS:10.72 | ZC:7 | CZC:9 | CE:249.25 | e\_ss:0.00 | isStb:True \\
\#8/20 | Type:FSF | K1:7.80 | K2:7.50 | K3:35.00 | K4:6.80 | MSE:0.0501 | Ts:3.86 | Tr:1.96 | \%OS:12.20 | ZC:8 | CZC:9 | CE:252.79 | e\_ss:0.00 | isStb:True \\
\#9/20 | Type:FSF | K1:7.50 | K2:7.50 | K3:35.00 | K4:6.50 | MSE:0.0505 | Ts:3.94 | Tr:2.04 | \%OS:10.72 | ZC:7 | CZC:9 | CE:249.25 | e\_ss:0.00 | isStb:True \\
\#10/20 | Type:FSF | K1:6.00 | K2:9.00 | K3:40.00 | K4:7.00 | MSE:0.0591 | Ts:3.02 | Tr:3.11 | \%OS:1.75 | ZC:4 | CZC:9 | CE:176.17 | e\_ss:0.00 | isStb:True \\
\textbf{=== \faStop\ TERMINATOR DECISION: TERMINATE\_SUCCESS ===} \\
Reasoning: All target metrics (MSE, settling time, and overshoot) have been successfully met or exceeded. The system is stable with no oscillations, and the performance trends show consistent improvement. Although parameter convergence isn't complete, the controller's effectiveness justifies termination. \\
\textbf{\faCheck\ Scenario 3 completed successfully!} \\
\\
\textbf{=== FINAL REPORT ===} \\
Completed 3 of 3 scenarios \\
\textbf{\faTrophy\ All scenarios completed successfully!}
\end{systemmetricsbox}

\end{document}